\documentclass[useAMS,usenatbib]{mn2e}
\usepackage{graphicx}

\title[Eclipsing PCEBs]{Orbital Period Variations in Eclipsing Post Common Envelope Binaries}

\author[S. G. Parsons et al.]{S.~G.~Parsons$^{1}$\thanks{steven.parsons@warwick.ac.uk},
T.~R.~Marsh$^{1}$,
C.~M.~Copperwheat$^{1}$,
V.~S.~Dhillon$^{2}$,
\newauthor
S.~P.~Littlefair$^{2}$
R.~D.~G.~Hickman$^{1}$
P.~F.~L.~Maxted$^{3}$
B.~T.~G{\"a}nsicke$^{1}$
\newauthor
E.~Unda-Sanzana$^{4}$ 
J.~P.~Colque$^{4}$
N.~Barraza$^{4}$
N.~S{\'a}nchez$^{4}$
and L.~A.~G.~Monard$^{5}$ \\
$^{1}$Department of Physics, University of Warwick, Coventry, CV4 7AL \\
$^{2}$Department of Physics and Astronomy, University of Sheffield, Sheffield S3 7RH\\
$^{3}$Astrophysics Group, Keele University, Keele, Staffodshire ST5 5BG\\
$^{4}$Instituto de Astronom{\'i}a, Universidad Cat{\'o}lica del Norte, Avenida Angamos 0610, Antofagasta, Chile\\
$^{5}$Bronberg Observatory, CBA Pretoria, P.O. Box 11426, Tiegerpoort 0056, South Africa}
\begin{document}
\input{pjw_aas_macros.cls}
\date{Accepted 2010 May 20. Received 2010 May 20; in original form 2010 March 23}

\pagerange{\pageref{firstpage}--\pageref{lastpage}} \pubyear{2010}

\maketitle

\label{firstpage}

\begin{abstract}
We present high speed ULTRACAM photometry of the eclipsing post common envelope binaries DE CVn, GK Vir, NN Ser, QS Vir, RR Cae, RX J$2130.6+4710$, SDSS $0110+1326$ and SDSS $0303+0054$ and use these data to measure precise mid-eclipse times in order to detect any period variations. We detect a large ($\sim 250$ sec) departure from linearity in the eclipse times of QS Vir which Applegate's mechanism fails to reproduce by an order of magnitude. The only mechanism able to drive this period change is a third body in a highly elliptical orbit. However, the planetary/sub-stellar companion previously suggested to exist in this system is ruled out by our data. Our eclipse times show that the period decrease detected in NN Ser is continuing, with magnetic braking or a third body the only mechanisms able to explain this change. The planetary/sub-stellar companion previously suggested to exist in NN Ser is also ruled out by our data. Our precise eclipse times also lead to improved ephemerides for DE CVn and GK Vir. The width of a primary eclipse is directly related to the size of the secondary star and variations in the size of this star could be an indication of Applegate's mechanism or Wilson (starspot) depressions which can cause jitter in the O-C curves. We measure the width of primary eclipses for the systems NN Ser and GK Vir over several years but find no definitive variations in the radii of the secondary stars. However, our data are precise enough ($\Delta R_\mathrm{sec} / R_\mathrm{sec} < 10^{-5}$) to show the effects of Applegate's mechanism in the future. We find no evidence of Wilson depressions in either system. We also find tentative indications that flaring rates of the secondary stars depend on their mass rather than rotation rates.

\end{abstract}

\begin{keywords}
binaries: eclipsing -- stars: evolution -- stars: late-type -- white dwarfs -- planetary systems
\end{keywords}

\section{Introduction}

Angular momentum loss drives the evolution of close binary stars. For short period systems ($< 3$ hours) gravitational radiation (\citealt{kraft62}; \citealt{faulkner71}) dominates whilst for longer period systems ($> 3$ hours) a magnetised stellar wind can extract angular momentum, the so called magnetic braking mechanism (\citealt{verbunt81}; \citealt{rappaport83}).

In the magnetic braking mechanism, charged particles from the main sequence star are trapped within its magnetised wind and forced to co-rotate with it. By dragging these particles around, the star transfers angular momentum to them slowing down its rotation. In close binaries the rotational and orbital periods have become synchronised meaning that the angular momentum is taken from the binary orbit causing the period to decrease. In the disrupted magnetic braking mechanism this process ceases in very low mass stars ($M\la0.3 M_{\sun}$) since they become fully convective and the magnetic field is no longer rooted to the stellar core. One of the great successes of this model is that it can explain the cataclysmic variable period gap (a dearth of systems with periods between 2 and 3 hours) since at periods of around 3 hours the secondary star becomes fully convective and shrinks back to within its Roche lobe stopping mass transfer. During this time the period loss is driven solely by gravitational radiation until the secondary star touches its Roche lobe again at a period of around 2 hours. However, it is still unclear how the angular momentum loss changes over the fully convective boundary \citep{andronov03}.

Accurate eclipse timings can reveal period changes; long term period decreases are the result of angular momentum loss, however, shorter timescale period modulation can be the result of Applegate's mechanism \citep{applegate92} or possible light travel time effects caused by the presence of a third body. In Applegate's mechanism, as one or both component stars go through activity cycles, the outer parts of the stars are subject to a magnetic torque changing the distribution of angular momentum and thus their oblateness. The orbit of the stars are gravitationally coupled to variations in their shape hence the period is altered on the same timescale as the activity cycles. This has the effect of modulating the period with fairly large amplitudes ($\Delta P/P \sim 10^{-5}$) over timescales of decades or longer.

The presence of a third body results in the central binary being displaced over the orbital period of the third body. This delays or advances eclipse times through variations in light travel time. Since the third body can have a large range of orbital periods these effects can happen over a range of timescales. Therefore, accurate eclipse timings of binaries can test theories of angular momentum loss as well as theories of stellar structure and potentially identify low mass companions.

When the more massive member of a binary evolves off the main-sequence it may, depending upon the orbital separation, fill its Roche lobe on either the giant or asymptotic giant branches. This can initiate a dynamically unstable mass transfer to the less massive component. If the latter is unable to accrete the material, a common envelope is formed containing the core of the giant and the companion star. Frictional forces within this envelope cause the two stars to spiral inwards. The ensuing loss of angular momentum expels the common envelope revealing the tightly bound core and companion. The resulting system is known as a Post Common Envelope Binary (PCEB). A small number of these systems are inclined in such a way that, as viewed from Earth, they exhibit eclipses, as the main sequence secondary star passes in front of the white dwarf primary. These deep eclipses allow us to measure precise mid-eclipse times and therefore detect any period variations. PCEBs have the added advantage that they are relatively simple systems and therefore accurate system parameters can be obtained helping to further constrain the mechanisms able to produce period changes.

Here we present high speed ULTRACAM photometry of eight eclipsing PCEBs and use these data to determine accurate and precise mid-eclipse times. We combine these with previous eclipse times to analyse any period variations in these systems.

\section{Observations}

\begin{table*}
 \centering
  \caption{ULTRACAM observations of eclipsing PCEBs. ``Av exp time'' is the average exposure time in seconds. Primary eclipses occur at phase 1, 2 etc.}
  \label{obs_jour1}
  \begin{tabular}{@{}lccccccccl@{}}
  \hline
  Date at      &Target          &Filters    &Telescope &UT       &UT     &Av exp   &Phase  &Comp & Conditions             \\
  start of run &                &           &          &start    &end    &time (s) &range  &star & (Transparency, seeing) \\
 \hline
 17/05/2002 & NN Ser         & \emph{u'g'r'} & WHT & 21:54:40 & 02:07:54 & 2.4 & 0.85--2.13 & 3      & Good, $\sim$1.2 arcsec \\
 18/05/2002 & RX J2130+4710  & \emph{u'g'r'} & WHT & 02:29:55 & 05:48:36 & 1.1 & 0.97--1.23 & 1      & Good, $\sim$1.2 arcsec \\
 18/05/2002 & NN Ser         & \emph{u'g'r'} & WHT & 21:21:20 & 02:13:17 & 3.9 & 0.39--1.23 & 2      & Variable, 1.2-2.4 arcsec \\
 19/05/2002 & GK Vir         & \emph{u'g'r'} & WHT & 21:09:08 & 23:58:00 & 2.1 & 0.89--1.02 & 1      & Good, $\sim$1.5 arcsec \\
 19/05/2002 & NN Ser         & \emph{u'g'r'} & WHT & 23:58:22 & 00:50:52 & 2.0 & 0.93--1.10 & 2      & Fair, $\sim$2 arcsec \\
 20/05/2002 & QS Vir         & \emph{u'g'r'} & WHT & 20:51:44 & 00:31:07 & 1.1 & 0.48--1.55 & 1      & Fair, $\sim$2 arcsec \\
 21/05/2002 & NN Ser         & \emph{u'g'r'} & WHT & 00:58:23 & 01:57:18 & 2.3 & 0.87--1.14 & 2      & Fair, $\sim$2 arcsec \\
 19/05/2003 & NN Ser         & \emph{u'g'z'} & WHT & 22:25:33 & 01:02:25 & 6.7 & 0.47--1.12 & 2      & Variable, 1.5-3 arcsec \\
 20/05/2003 & QS Vir         & \emph{u'g'i'} & WHT & 23:43:55 & 00:53:24 & 2.9 & 0.93--1.64 & 2      & Variable, 1.2-3 arcsec \\
 22/05/2003 & NN Ser         & \emph{u'g'i'} & WHT & 00:29:00 & 04:27:32 & 1.9 & 0.32--1.59 & 2      & Excellent, $\sim$1 arcsec \\
 22/05/2003 & DE CVn         & \emph{u'g'i'} & WHT & 21:57:16 & 22:22:44 & 1.4 & 0.15--0.18 & 1,2    & Good, $\sim$1.5 arcsec \\
 22/05/2003 & GK Vir         & \emph{u'g'i'} & WHT & 23:25:42 & 00:38:23 & 5.0 & 0.92--1.05 & 1,2    & Excellent, $<$1 arcsec \\
 23/05/2003 & QS Vir         & \emph{u'g'i'} & WHT & 00:39:32 & 01:35:03 & 2.9 & 0.84--1.10 & 2      & Good, $\sim$1.5 arcsec \\
 23/05/2003 & NN Ser         & \emph{u'g'i'} & WHT & 03:24:57 & 03:50:40 & 2.0 & 0.37--1.08 & 2      & Excellent, $<$1 arcsec \\
 23/05/2003 & GK Vir         & \emph{u'g'i'} & WHT & 00:31:54 & 01:22:28 & 4.0 & 0.95--1.04 & 1,2    & Excellent, $<$1 arcsec \\ 
 24/05/2003 & GK Vir         & \emph{u'g'i'} & WHT & 20:51:43 & 22:00:41 & 4.0 & 0.40--0.54 & 1,2    & Good, $\sim$1.2 arcsec \\
 24/05/2003 & QS Vir         & \emph{u'g'i'} & WHT & 22:02:51 & 22:56:24 & 2.9 & 0.38--1.07 & 2      & Good, $\sim$1.2 arcsec \\
 24/05/2003 & NN Ser         & \emph{u'g'i'} & WHT & 22:58:55 & 23:33:49 & 2.0 & 0.90--1.09 & 2      & Good, $\sim$1.2 arcsec \\
 25/05/2003 & DE CVn         & \emph{u'g'i'} & WHT & 00:34:21 & 01:44:03 & 1.4 & 0.94--1.06 & 1,2    & Good, $\sim$1.2 arcsec \\
 25/05/2003 & RX J2130+4710  & \emph{u'g'i'} & WHT & 02:41:38 & 03:48:08 & 1.4 & 0.96--1.03 & 2      & Excellent, $\sim$1 arcsec \\
 25/05/2003 & DE CVn         & \emph{u'g'i'} & WHT & 22:33:00 & 23:59:13 & 1.4 & 0.44--0.60 & 1,2    & Good, $\sim$1.2 arcsec \\
 26/05/2003 & NN Ser         & \emph{u'g'i'} & WHT & 01:29:45 & 02:15:58 & 2.0 & 0.39--0.64 & 2      & Excellent, $\sim$1 arcsec \\
 26/05/2003 & RX J2130+4710  & \emph{u'g'i'} & WHT & 03:41:26 & 04:43:31 & 1.4 & 0.95--1.03 & 2      & Good, $\sim$1.5 arcsec \\
 13/11/2003 & RX J2130+4710  & \emph{u'g'r'} & WHT & 19:04:41 & 21:44:06 & 1.1 & 0.39--0.59 & 2      & Good, $\sim$1.5 arcsec \\
 03/05/2004 & NN Ser         & \emph{u'g'i'} & WHT & 22:13:44 & 05:43:11 & 2.5 & 0.95--3.27 & 2      & Variable, 1.2-3.2 arcsec \\
 04/05/2004 & DE CVn         & \emph{u'g'i'} & WHT & 20:39:25 & 22:56:47 & 0.6 & 0.68--0.96 & 1,2    & Excellent, $\sim$1 arcsec \\ 
 04/05/2004 & NN Ser         & \emph{u'g'i'} & WHT & 23:18:46 & 23:56:59 & 2.5 & 0.90--1.61 & 2      & Variable, 1.2-3 arcsec \\
 05/05/2004 & GK Vir         & \emph{u'g'i'} & WHT & 01:36:55 & 04:01:29 & 3.9 & 0.84--1.12 & 1,2    & Excellent, $\sim$1 arcsec \\ 
 25/11/2005 & RR Cae         & \emph{u'g'i'} & VLT & 00:21:31 & 01:22:04 & 0.5 & 0.42--0.56 & 1      & Good, $\sim$1.5 arcsec \\
 26/11/2005 & RR Cae         & \emph{u'g'i'} & VLT & 23:53:01 & 00:44:38 & 0.5 & 0.97--1.06 & 1      & Good, $\sim$1.5 arcsec \\
 27/11/2005 & RR Cae         & \emph{u'g'i'} & VLT & 07:04:42 & 08:16:30 & 0.5 & 0.93--1.10 & 1      & Good, $\sim$1.5 arcsec \\
 09/03/2006 & DE CVn         & \emph{u'g'r'} & WHT & 23:08:36 & 01:00:23 & 1.4 & 0.90--1.10 & 1,2    & Poor, $>$3 arcsec \\
 10/03/2006 & NN Ser         & \emph{u'g'r'} & WHT & 01:02:34 & 06:46:49 & 2.0 & 0.91--2.70 & 2      & Variable, 1.2-3 arcsec \\
 11/03/2006 & GK Vir         & \emph{u'g'r'} & WHT & 00:04:21 & 01:06:29 & 3.0 & 0.96--1.08 & 1,2    & Variable, 1-3 arcsec \\
 11/03/2006 & DE CVn         & \emph{u'g'r'} & WHT & 01:35:20 & 03:00:41 & 1.2 & 0.92--1.08 & 1,2    & Variable, 1-3 arcsec \\
 11/03/2006 & GK Vir         & \emph{u'g'r'} & WHT & 04:00:04 & 04:56:25 & 3.0 & 0.43--0.55 & 1,2    & Variable, 1-3 arcsec \\
 11/03/2006 & NN Ser         & \emph{u'g'r'} & WHT & 05:01:13 & 05:50:14 & 2.0 & 0.85--1.11 & 2      & Excellent, $<$1 arcsec \\
 12/03/2006 & GK Vir         & \emph{u'g'r'} & WHT & 00:35:43 & 01:39:36 & 3.0 & 0.93--1.06 & 1,2    & Excellent, $\sim$1 arcsec \\
 12/03/2006 & DE CVn         & \emph{u'g'r'} & WHT & 03:50:25 & 05:06:55 & 1.2 & 0.92--1.07 & 1,2    & Good, $\sim$1.2 arcsec \\
 12/03/2006 & DE CVn         & \emph{u'g'r'} & WHT & 21:40:15 & 22:30:25 & 1.2 & 0.96--1.06 & 1,2    & Poor, $>$3 arcsec \\
 13/03/2006 & QS Vir         & \emph{u'g'r'} & WHT & 00:42:35 & 01:34:29 & 2.4 & 0.88--1.09 & 2      & Fair, $\sim$2 arcsec \\
 13/03/2006 & GK Vir         & \emph{u'g'r'} & WHT & 01:38:42 & 02:20:03 & 3.0 & 0.96--0.99 & 1,2    & Poor, $>$3 arcsec \\
 10/06/2007 & NN Ser         & \emph{u'g'i'} & VLT & 04:59:25 & 05:46:18 & 0.9 & 0.40--0.61 & 1,2    & Excellent, $\sim$1 arcsec \\
 17/06/2007 & NN Ser         & \emph{u'g'i'} & VLT & 03:57:48 & 04:54:39 & 2.0 & 0.86--1.14 & 1,2    & Good, $\sim$1.2 arcsec \\
 18/06/2007 & NN Ser         & \emph{u'g'i'} & VLT & 01:50:16 & 02:38:09 & 1.0 & 0.86--1.10 & 1,2    & Excellent, $<$1 arcsec \\
 18/06/2007 & GK Vir         & \emph{u'g'i'} & VLT & 02:40:17 & 05:24:46 & 1.5 & 0.81--1.14 & 3      & Excellent, $<$1 arcsec \\
 17/10/2007 & SDSS 0303+0054 & \emph{u'g'i'} & WHT & 02:25:40 & 03:31:11 & 5.0 & 0.89--1.18 & 1,2    & Good, $\sim$1.2 arcsec \\
 18/10/2007 & SDSS 0303+0054 & \emph{u'g'i'} & WHT & 02:25:04 & 06:25:18 & 5.2 & 0.28--1.52 & 1,2    & Good, $\sim$1.2 arcsec \\
 21/10/2007 & SDSS 0110+1326 & \emph{u'g'i'} & WHT & 02:46:50 & 04:32:04 & 1.2 & 0.86--1.07 & 1      & Good, $\sim$1.2 arcsec \\
 29/10/2007 & SDSS 0303+0054 & \emph{u'g'i'} & WHT & 04:40:07 & 05:36:14 & 2.3 & 0.80--1.09 & 1,2    & Poor, $>$3 arcsec \\
 07/08/2008 & NN Ser         & \emph{u'g'r'} & WHT & 23:41:29 & 00:22:46 & 2.8 & 0.86--1.07 & 2      & Excellent, $<$1 arcsec \\
\hline
\end{tabular}
\end{table*}

\begin{table*}
 \centering
 \begin{minipage}{\textwidth}
  \centering
  \caption{Other observations of eclipsing PCEBs. Primary eclipses occur at phase 1, 2 etc.}
  \label{obs_jour2}
  \begin{tabular}{@{}lcccccccl@{}}
  \hline
  Date at start &Target  &Filter/     &Obs\footnote{CBA: Bronberg Observatory, Pretoria, South Africa. ESO: European Southern Observatory 3.6m telescope, La Silla, Chile. NTT: New Technology Telescope, La Silla, Chile. OCA: Observatorio Cerro Armazones, Chile.} &UT        &UT      &Average     &Phase  &Conditions             \\
  of run    &        &Instrument  &           &start     &end     &exp time (s)&range  &(Transparency, seeing) \\
 \hline
 10/07/2006 & QS Vir & Unfiltered & CBA       & 17:29:27 & 18:40:50 & 30.0 & 0.80--1.14 & Clear \\
 11/07/2006 & QS Vir & Unfiltered & CBA       & 18:52:22 & 19:57:09 & 30.0 & 0.83--1.12 & Clear \\
 13/07/2006 & QS Vir & Unfiltered & CBA       & 18:01:30 & 18:48:48 & 30.0 & 0.86--1.07 & Clear \\
 18/07/2006 & QS Vir & V          & CBA       & 17:19:32 & 18:26:58 & 30.0 & 0.83--1.14 & Clear \\
 19/07/2006 & QS Vir & V          & CBA       & 18:26:42 & 19:44:02 & 30.0 & 0.77--1.12 & Clear \\
 20/07/2006 & QS Vir & I          & CBA       & 16:24:46 & 17:28:17 & 30.0 & 0.84--1.13 & Clear \\
 23/07/2006 & QS Vir & Unfiltered & CBA       & 16:18:44 & 20:24:22 & 30.0 & 0.71--1.84 & Cloudy \\
 29/07/2006 & QS Vir & Unfiltered & CBA       & 17:27:15 & 18:35:11 & 30.0 & 0.82--1.13 & Clear \\
 06/02/2008 & QS Vir & ULTRASPEC  & ESO       & 07:43:05 & 08:48:32 & 0.4  & 0.90--1.07 & Excellent, $\sim$1 arcsec \\
 07/02/2008 & QS Vir & ULTRASPEC  & ESO       & 05:29:42 & 06:41:22 & 2.0  & 0.91--1.17 & Good, $\sim$1.5 arcsec \\
 08/02/2008 & QS Vir & ULTRASPEC  & ESO       & 06:51:30 & 07:40:12 & 2.0  & 0.87--1.07 & Good, $\sim$1.5 arcsec \\
 09/02/2008 & QS Vir & ULTRASPEC  & ESO       & 08:16:11 & 08:59:51 & 2.0  & 0.89--1.07 & Excellent, $\sim$1 arcsec \\
 10/06/2009 & NN Ser & ULTRASPEC  & NTT       & 04:48:15 & 04:44:15 & 1.9  & 0.91--1.06 & Good, $\sim$1.5 arcsec \\
 27/01/2010 & QS Vir & Unfiltered & OCA       & 06:22:25 & 07:58:31 & 10.0 & 0.97--1.41 & Clear \\
 28/01/2010 & QS Vir & Unfiltered & OCA       & 06:02:14 & 09:34:46 & 10.0 & 0.51--1.49 & Clear \\
 30/01/2010 & QS Vir & Unfiltered & OCA       & 06:55:35 & 08:56:50 & 10.0 & 0.02--0.58 & Cloudy \\
 31/01/2010 & QS Vir & Unfiltered & OCA       & 06:22:00 & 09:43:21 & 10.0 & 0.50--1.43 & Clear \\
 07/02/2010 & QS Vir & Unfiltered & OCA       & 05:40:58 & 09:11:54 & 10.0 & 0.75--1.72 & Cloudy \\
 08/02/2010 & QS Vir & Unfiltered & OCA       & 06:07:58 & 09:44:52 & 10.0 & 0.51--1.51 & Clear \\
\hline
\end{tabular}
\end{minipage}
\end{table*}

ULTRACAM is a high-speed, triple-beam CCD camera \citep{dhillon07} which can acquire simultaneous readings in the SDSS \emph{u'} and \emph{g'} filters and either \emph{r'}, \emph{i'} or \emph{z'} filters. Most of our observations use the \emph{i'} filter in the red arm but on a number of occasions the \emph{r'} filter was used instead. The \emph{z'} filter was used once in 2003 May. The data were collected with ULTRACAM mounted as a visitor instrument on the 4.2-m William Herschel Telescope (WHT) or the 8.2-m Very Large Telescope (VLT). A complete log of all ULTRACAM observations of eclipsing PCEBs is given in Table~\ref{obs_jour1}. 

In addition to the ULTRACAM data, we obtained photometry of QS Vir using the Meade 12.5 inch telescope at Bronberg Observatory, Pretoria and the 0.84-m telescope at the Observatorio Cerro Armazones using a SBIG ST-10 camera. We also measure eclipse times from ULTRASPEC \citep{dhillon08} observations of QS Vir and NN Ser. Table~\ref{obs_jour2} summerises all non-ULTRACAM observations.

For the ULTRACAM data, we windowed the CCD in order to achieve exposure times of 2-3 s which we varied to account for the conditions, with the exception of RX J$2130.6+4710$ for which we used shorter exposure times since it lies only 12 arcsec away from a bright star (HD 204906, $V=8.45$). We also used shorter exposure times for the bright target DE CVn. The dead time between exposures was $\sim 25$ ms.

All of these data were reduced using the ULTRACAM pipeline software. Debiassing, flatfielding and sky background subtraction were performed in the standard way. The source flux was determined with aperture photometry using a variable aperture, whereby the radius of the aperture is scaled according to the FWHM. Variations in observing conditions were accounted for by determining the flux relative to a comparison star in the field of view. The comparison stars used in each run are also listed in Table~\ref{obs_jour1}. Apparent magnitudes and coordinates for each of the comparison stars used are given in Table~\ref{comps}. As already mentioned RX J$2130.6+4710$ lies close to a bright source and in order to correct for this we used the same procedure as in \citet{maxted04} whereby another aperture was placed on the sky at the same distance from the bright star as RX J$2130.6+4710$ and symmetrically located with respect to the diffraction spikes from the bright star. This was used to correct for scattered light from the bright star. We flux calibrated our targets by determining atmospheric extinction coefficients in each of the bands in which we observed and calculated the absolute flux of our targets using observations of standard stars (from \citealt{Smith02}) taken in twilight. Using our absorption coefficients we extrapolated all fluxes to an airmass of $0$. The systematic error introduced by our flux calibration is  $<0.1$ mag in all bands.

\begin{table}
 \centering
  \caption{Comparison star apparent magnitudes and offsets from the targets.}
  \label{comps}
  \begin{tabular}{@{}lcccccc@{}}
  \hline
Star & \emph{u'} & \emph{g'} & \emph{r'} & \emph{i'} & RA        & DEC     \\
     &           &           &           &           &           &         \\
\hline
\multicolumn{4}{|l|}{\bf{DE CVn:}} \\
1    & 15.9      & 13.4      & 12.5      & 12.1      & 13:26:28.10 & +45:33:11.5 \\
2    & 15.3      & 13.9      & 13.4      & 13.3      & 13:26:39.00 & +45:34:54.1 \\
\multicolumn{4}{|l|}{\bf{GK Vir:}} \\
1    & 16.5      & 15.2      & 14.8      & 14.6      & 14:15:31.94 & +01:16:35.8 \\
2    & 16.7      & 15.2      & 14.6      & 14.5      & 14:15:32.93 & +01:21:04.9 \\
3    & 16.6      & 15.1      & --        & 14.4      & 14:15:35.96 & +01:19:42.3 \\
\multicolumn{4}{|l|}{\bf{NN Ser:}} \\
1    & 17.0      & 15.6      & 15.8      & 15.0      & 15:52:53.82 & +12:54:45.8 \\
2    & 14.6      & 13.4      & 13.7      & 12.8      & 15:52:48.22 & +12:56:27.5 \\
3    & 16.7      & 14.6      & 13.7      & --        & 15:52:54.66 & +12:53:11.2 \\
\multicolumn{4}{|l|}{\bf{QS Vir:}} \\
1    & 14.0      & 11.3      & 10.2      & --        & 13:49:37.44 & -13:10:25.3 \\
2    & 15.4      & 13.7      & 13.4      & 12.9      & 13:49:51.79 & -13:10:58.0 \\
\multicolumn{4}{|l|}{\bf{RR Cae:}} \\
1    & 18.8      & 17.0      & --        & 16.1      & 04:21:10.44 & -48:37:24.6 \\
\multicolumn{4}{|l|}{\bf{RX J2130+4710:}}\\
1    & 14.5      & 12.2      & 12.2      & --        & 21:30:19.69 & +47:10:26.0 \\
2    & 13.9      & 12.7      & 12.6      & 11.9      & 21:30:12.20 & +47:10:39.8 \\
\multicolumn{4}{|l|}{\bf{SDSS 0110+1326:}}\\
1    & 13.7      & 12.6      & --        & 12.3      & 01:10:01.58 & +13:28:33.1 \\
\multicolumn{4}{|l|}{\bf{SDSS 0303+0054:}}\\
1    & 17.4      & 16.3      & --        & 15.6      & 03:03:11.74 & +00:54:58.5 \\
2    & 17.8      & 15.5      & --        & 13.3      & 03:03:11.16 & +00:54:03.1 \\
\hline
\end{tabular}
\end{table}

\begin{figure*}
\begin{center}
 \includegraphics[width=0.99\textwidth]{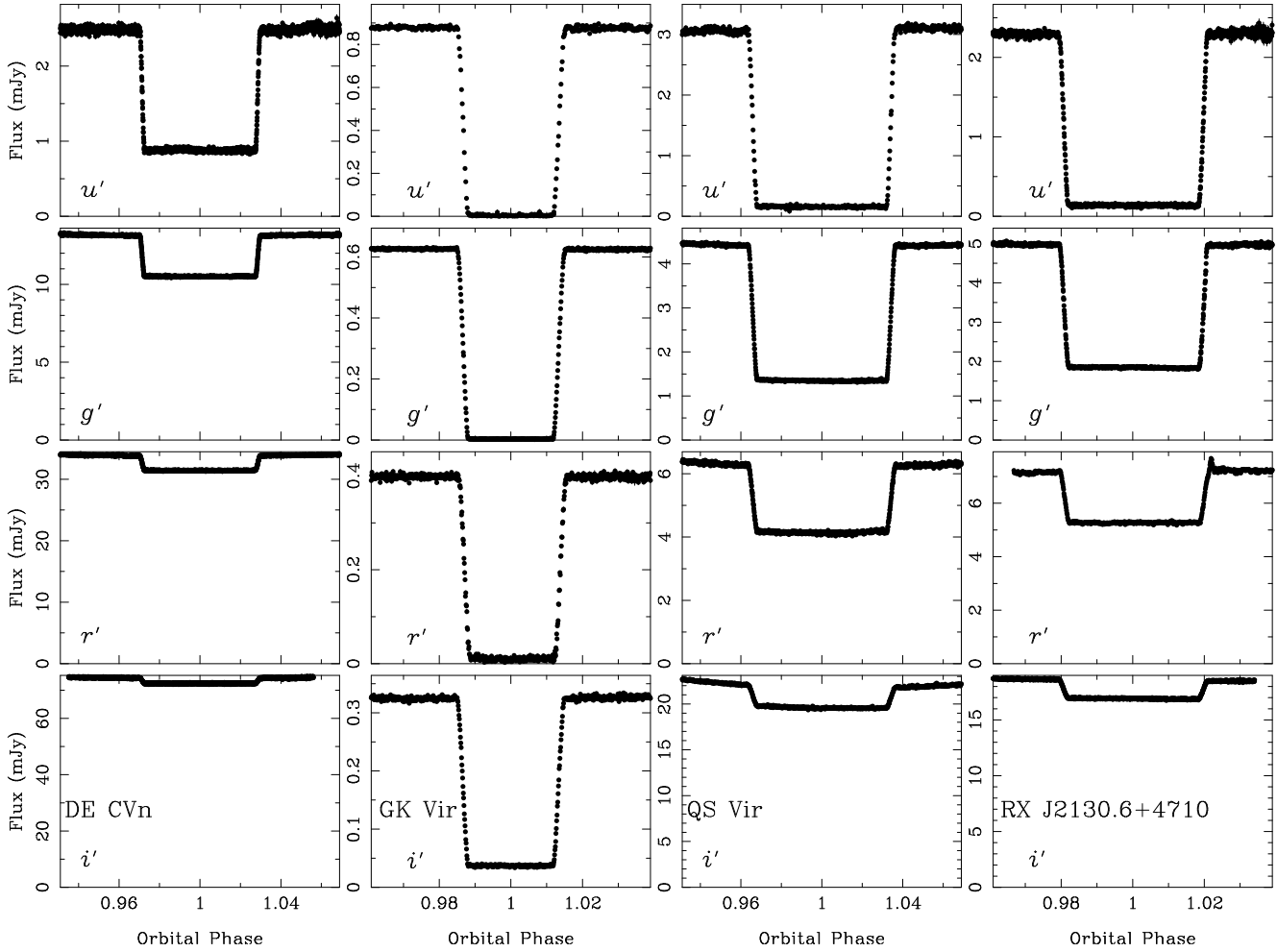}
 \caption{Flux calibrated primary eclipses of (\emph{left} to \emph{right}) DE CVn, GK Vir, QS Vir and RX J$2130.6+4710$ in (\emph{top} to \emph{bottom}) \emph{u'} band, \emph{g'} band, \emph{r'} band and  \emph{i'} band. Light curves were made by phase binning all available eclipses then combining them. Any flares were removed before the light curves were combined with the exception of RX J$2130.6+4710$ in the \emph{r'} band where there was only one eclipse which featured a flare.}
 \label{priecl1}
\end{center}
\end{figure*}

\begin{figure*}
\begin{center}
 \includegraphics[width=0.95\textwidth]{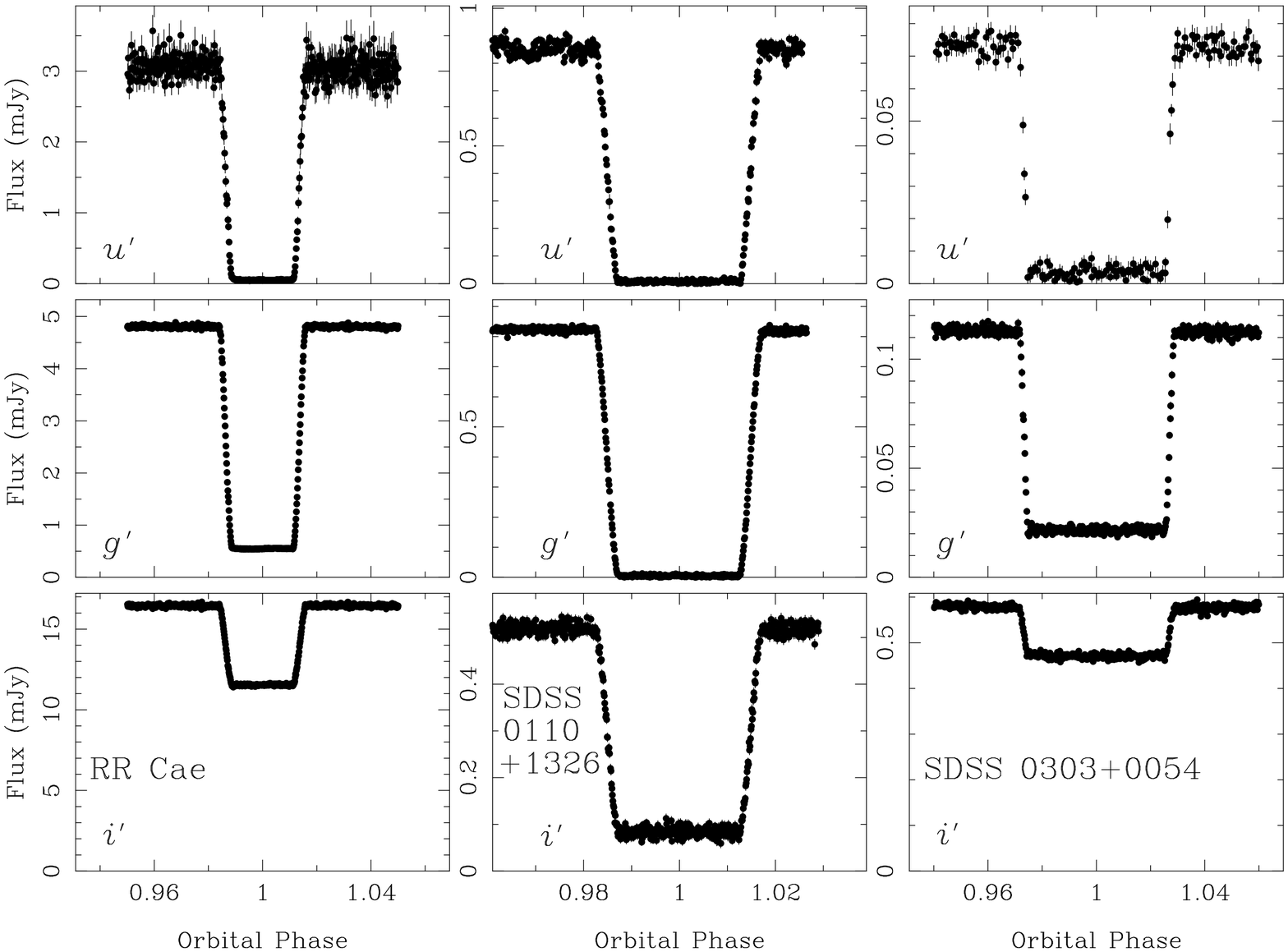}
 \caption{Flux calibrated primary eclipses of (\emph{left} to \emph{right}) RR Cae, SDSS $0110+1326$ and SDSS $0303+0054$ in (\emph{top} to \emph{bottom}) \emph{u'} band, \emph{g'} band and \emph{i'} band. Light curves were made by phase binning all available eclipses then combining them. Any flares were removed before the light curves were combined. Longer exposures were used in the \emph{u'} band for SDSS $0303+0054$ since it is very faint in this band. A micro-flare occurs during the ingress of the SDSS $0110+1326$ eclipse, visible in the \emph{u'} band light curve.}
 \label{priecl2}
\end{center}
\end{figure*}

To correct for extinction differences between our targets and the comparison star we determined the comparison star colours using the same method described above, then determined the colour dependent difference in extinction coefficients for the comparison star and the target using a theoretical extinction vs. colour plot\footnote{theoretical extinction vs. colour plots for ULTRACAM are available at http://garagos.net/dev/ultracam/filters}. The additional extinction correction is then given by
\begin{eqnarray}
10^{\left(k_{T} - k_{C}\right)X/2.5},
\end{eqnarray}
where k$_{T}$ is the extinction coefficient for the target, k$_{C}$ is the extinction coefficient for the comparison and X is the airmass.

\section{Light Curves}

We phase binned all ULTRACAM data for each target using published ephemerides. For light curves with primary eclipses we calculated the observed minus calculated (O-C) eclipse times (see Section~\ref{OCtimes} for eclipse timings) and adjusted the phase of the light curve accordingly. For those light curves with no primary eclipse, we extrapolated the phase correction from nearby O-C times.

For a given target, data within each phase bin were averaged using inverse variance weights whereby data with smaller errors are given larger weightings; we removed any data affected by flares (see Section~\ref{flaresec}). This results in a set of light curves for each target in each band observed. There are \emph{u'}, \emph{g'} and \emph{i'} data for all targets but several have not been observed in the \emph{r'} band. Figure~\ref{priecl1} shows the primary eclipses of those systems observed in all four bands, Figure~\ref{priecl2} shows the primary eclipses of those systems observed in the \emph{u'}, \emph{g'} and \emph{i'} bands. We also show the light curves of those systems with full orbital coverage.

\begin{table*}
 \centering
 \begin{minipage}{\textwidth}
  \centering
  \caption{Previously determined physical parameters for the eclipsing PCEBs observed with ULTRACAM. Out of ecl \emph{g'} is the average \emph{g'} band magnitude of the system out of the primary eclipse.}
  \label{systems}
  \begin{tabular}{@{}lccccccccl@{}}
  \hline
System&P$_\mathrm{orb}$ & Out of       & M$_\mathrm{WD}$ &R$_\mathrm{WD}$ &T$_\mathrm{eff,WD}$ &M$_\mathrm{sec}$ &R$_\mathrm{sec}$ &Sp2 &Ref. \\
      & (d)            & ecl \emph{g'}& $(M_{\sun})$    & $(R_{\sun})$   & (K)               & $(M_{\sun})$    & $(R_{\sun})$    &    &     \\
 \hline
DE CVn    &0.3641& 13.50 &$0.51_{-0.02}^{+0.06}$&$0.0136_{-0.0002}^{+0.0008}$&$8000\pm1000$ &$0.41\pm0.06$  &$0.37_{-0.007}^{+0.06}$&M3V       &(1) \\
GK Vir    &0.3443& 16.81 &$0.51\pm0.04$       &0.016                     &$48800\pm1200$&0.1            &0.15                 &M3--5     &(2) \\
NN Ser    &0.1301& 16.43 &$0.535\pm0.012$     &$0.0211\pm0.0002$         &$57000\pm3000$&$0.111\pm0.004$&$0.149\pm0.002$      &M$4\pm0.5$&(3) \\
QS Vir    &0.1508& 14.68 &$0.77\pm0.04$       &$0.011\pm0.01$            &$14220\pm300$ &$0.51\pm0.04$  &$0.42\pm0.02$        &M3.5--4   &(4) \\
RR Cae    &0.3037& 14.58 &$0.44\pm0.022$      &$0.015\pm0.0004$          &$7540\pm175$  &$0.183\pm0.013$&$0.188-0.23$         &M4        &(5) \\
RX J2130  &0.5210& 14.55 &$0.554\pm0.017$     &$0.0137\pm 0.0014$        &$18000\pm1000$&$0.555\pm0.023$&$0.534\pm0.017$      &M3.5--4   &(6) \\
SDSS 0110 &0.3327& 16.53 &$0.47\pm0.2$        &$0.0163-0.0175$           &$25900\pm427$ &$0.255-0.38$   &$0.262-0.36$         &M3--5     &(7) \\
SDSS 0303 &0.1344& 18.60 &$0.878-0.946$       &$0.0085-0.0093$           &$<8000$       &$0.224-0.282$  &$0.246-0.27$         &M4--5     &(7) \\
\hline
\end{tabular}
\begin{flushleft}
Ref.: 1 -- \citet{besselaar07}; 2 -- \citet{fulbright93}; 3 -- \citet{parsons10}; 4 -- \citet{odonoghue03}; 5 -- \citet{maxted07}; 6 -- \citet{maxted04}; 7 -- \citet{pyrzas09}.
\end{flushleft}
\end{minipage}
\end{table*}

All of our targets have been studied previously and their basic parameters have been determined. Table~\ref{systems} lists these general parameters. Here we briefly introduce each system and describe their light curves.

\subsubsection*{DE CVn}

DE CVn (RX J$1326.9+4532$) is a bright ($V=12.8$) eclipsing binary consisting of a cool DA white dwarf primary and a M3V red dwarf secondary. It was discovered as an X-ray source by ROSAT \citep{voges99}. The period and eclipse depth were first measured by \citet{robb97}. The most recent analysis of this system was carried out by \citet{besselaar07} who determined the system parameters by combining spectroscopic and photometric observations including ULTRACAM data. We include their ULTRACAM data here along with more recent observations.

Our observations of DE CVn focus on the primary eclipse. DE CVn displays large ellipsoidal modulation and regular flaring. Its primary eclipse is shown in Figure~\ref{priecl1}. The secondary star dominates towards the red, therefore the \emph{i'} band primary eclipse is very shallow. 

\subsubsection*{GK Vir}

GK Vir (PG $1413+015$) is a faint ($V=17.0$) relatively unstudied PCEB with a hot DAO white dwarf primary and a low mass secondary discovered by \citet{green78}. \citet{fulbright93} combined the photometry from \citet{green78} and high resolution spectroscopy to constrain the system parameters. There are no other published observations of this system.

7 primary eclipses of GK Vir have been observed with ULTRACAM since 2002. GK Vir shows a reflection effect with an amplitude of 0.03 mag in the \emph{u'}, 0.04 mag in the \emph{g'}, 0.05 mag in the \emph{r'} and 0.06 mag in the \emph{i'} band caused by reprocessed light from the hemisphere of the secondary star facing the white dwarf. No flares have ever been detected in the light curves of GK Vir. Its primary eclipse is shown in Figure~\ref{priecl1}.

\begin{figure}
\begin{center}
 \includegraphics[width=0.99\columnwidth]{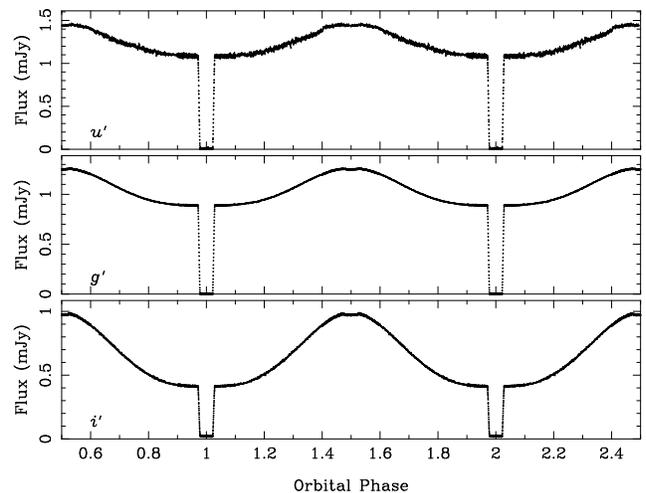}
 \caption{Full orbit light curves of NN Ser in (\emph{top} to \emph{bottom}) \emph{u'} band, \emph{g'} band and \emph{i'} band. NN Ser shows a large reflection effect. Smaller bins were used around both the primary and secondary eclipses.}
 \label{nnser_full}
\end{center}
\end{figure}

\subsubsection*{NN Ser}

NN Ser (PG $1550+131$) contains a hot DAO white dwarf primary with a low mass secondary. It was discovered in the Palomar Green Survey \citep{green82} and has been well studied. A period decrease in this system was seen by \citet{brinkworth06}. \citet{qian09} proposed that NN Ser has a sub-stellar companion based on eclipse timings. A thorough analysis of NN Ser was carried out by \citet{parsons10} using the ULTRACAM data presented here in combination with UVES spectroscopy. Here we look in detail at its eclipse times, and include additional data to those presented in \citet{parsons10}.

19 primary eclipses of NN Ser have been observed with ULTRACAM since 2002. Observations cover both the primary and secondary eclipses as well as some full orbit light curves. We have not detected any flaring events in over 37 hours of ULTRACAM photometry for NN Ser. \emph{u'}, \emph{g'} and \emph{i'} band full orbit light curves are shown in Figure~\ref{nnser_full}.

\begin{figure}
\begin{center}
 \includegraphics[width=0.99\columnwidth]{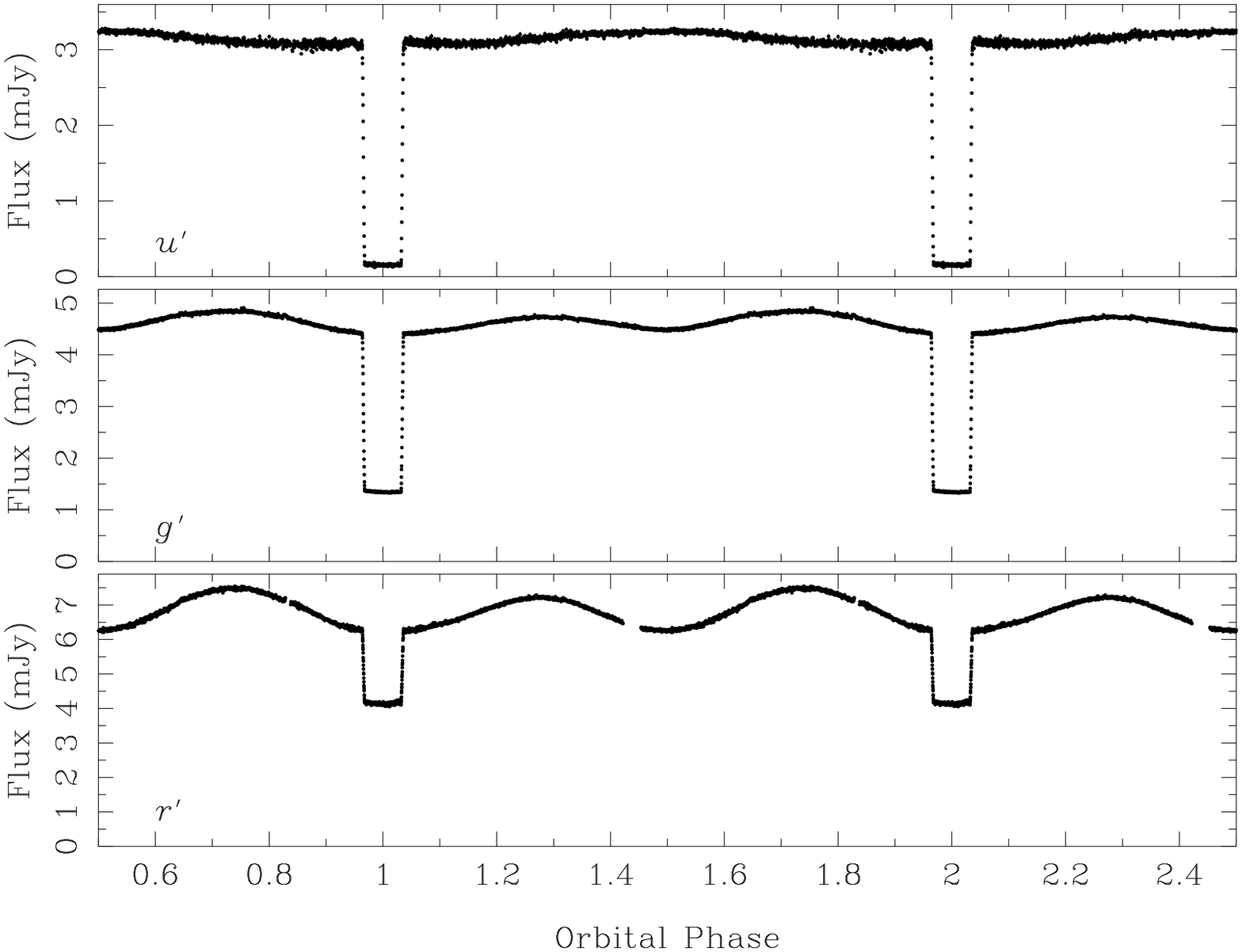}
 \caption{Full orbit light curves of QS Vir in (\emph{top} to \emph{bottom}) \emph{u'} band, \emph{g'} band and \emph{r'} band. A small reflection effect is visible in the \emph{u'} band light curve whilst ellipsoidal modulation is clearly seen in the \emph{g'} and \emph{r'} band light curves.}
 \label{qsvir_full}
\end{center}
\end{figure}

\subsubsection*{QS Vir}

QS Vir (EC $13471-1258$) was discovered in the Edinburgh-Cape faint blue object survey of high galactic latitudes \citep{kilkenny97}. The DA white dwarf primary has a companion red dwarf that is close to filling its Roche lobe \citep{odonoghue03}. \citet{odonoghue03} suggested that QS Vir is a hibernating cataclysmic variable. Recently \citet{qian10} proposed the existence of a giant planet in this system too by analysing the eclipse timings.

QS Vir was regularly observed with ULTRACAM between 2002 and 2006. Due to its short orbital period, QS Vir has full phase coverage. It shows regular flaring. The primary eclipse light curves are shown in Figure~\ref{priecl1}, the \emph{i'} band eclipse shows a clear gradient across the eclipse caused by the varying brightness of the secondary star across its surface. Figure~\ref{qsvir_full} shows full orbit light curves of QS Vir in the \emph{u'}, \emph{g'} and \emph{r'} bands. A small reflection effect is evident in the \emph{u'} band with an amplitude of 0.3 mag. Ellipsoidal modulation is evident in the \emph{g'} and \emph{r'} bands. 

\subsubsection*{RR Cae}

\begin{figure}
\begin{center}
 \includegraphics[width=0.99\columnwidth]{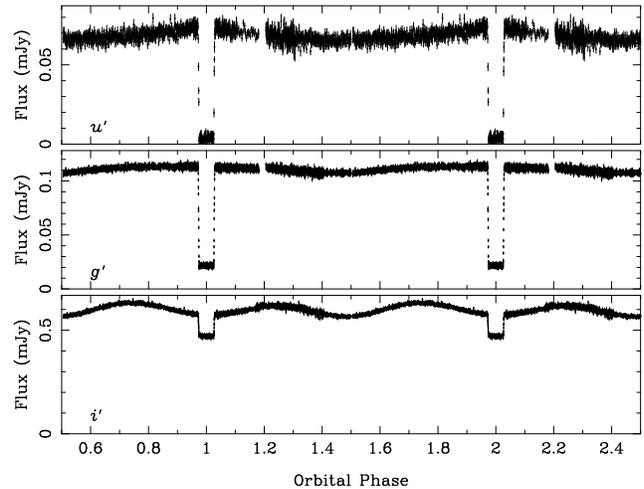}
 \caption{Full orbit light curves of SDSS $0303+0054$  in (\emph{top} to \emph{bottom}) \emph{u'} band, \emph{g'} band and \emph{i'} band. Ellipsoidal modulation is obvious in the \emph{i'} band.}
 \label{sd0303_full}
\end{center}
\end{figure}

Discovered as a high proper motion object by \citet{luyten55}, RR Cae contains a cool DA white dwarf with a low mass companion and has a deep primary eclipse \citep{krzeminski84}. Recently \citet{maxted07} used new photometry and spectroscopy to determine the mass and radius of the secondary star to high precision, they also noted a change in the shape of the primary eclipse from night to night. \citet{zuckerman03} detected spectacular metal absorption lines from the white dwarf.

RR Cae has only been observed with ULTRACAM twice, the first observation targeted the secondary eclipse, the second targeted the primary eclipse and recorded two. Unfortunately, RR Cae lacks any nearby bright comparison stars so a fairly dim comparison was used which adds some noise to the light curves in Figure~\ref{priecl2} particularly in the \emph{u'} band. RR Cae shows a high level of flaring with flares visible in each of the observations. 

\subsubsection*{RX J2130.6+4710}

RX J$2130.6+4710$ was discovered as a soft X-ray source by the \emph{ROSAT} satellite \citep{trumper82}; it contains a DA white dwarf primary. \citet{maxted04} used medium-resolution spectroscopy and ULTRACAM photometry to determine the system parameters. There have been no subsequent observations of RX J$2130.6+4710$ with ULTRACAM. However, we include their data here for completeness.
    
RX J$2130.6+4710$ was observed with ULTRACAM in 2002 and 2003. Three primary eclipses were observed. RX J$2130.6+4710$ lies only 12 arcsec away from a bright G0 star (HD 204906) making photometric extraction difficult. RX J$2130.6+4710$ shows high levels of flaring. The primary eclipse is shown in Figure~\ref{priecl1} in each band. The \emph{i} band light curve shows a gradient across the eclipse, caused by non-uniform surface brightness over the surface of the secondary star.

\subsubsection*{SDSS J011009.09+132616.1}

SDSS J$011009.09+132616.1$ (WD $0107+131$, henceforth SDSS $0110+1326$) was identified as an eclipsing post common envelope binary from the Sloan Digital Sky Survey (SDSS) by \citet{pyrzas09}. It contains a DA white dwarf with an M3--M5 companion.

Only one observation of SDSS $0110+1326$ with ULTRACAM exists. It targeted the primary eclipse as seen in Figure~\ref{priecl2}. There is a small flare on the ingress visible in the \emph{u'} band light curve, it is also present in the \emph{g'} band light curve though not visible in Figure~\ref{priecl2}.

\subsubsection*{SDSS J030308.35+005444.1}

SDSS J$030308.35+005444.1$ (SDSS J$030308.36+005443.7$ on SIMBAD, henceforth SDSS $0303+0054$) was also identified as an eclipsing PCEB by \citet{pyrzas09}. The DC white dwarf is the most massive white dwarf currently known in an eclipsing PCEB.

\begin{figure}
\begin{center}
 \includegraphics[width=0.99\columnwidth]{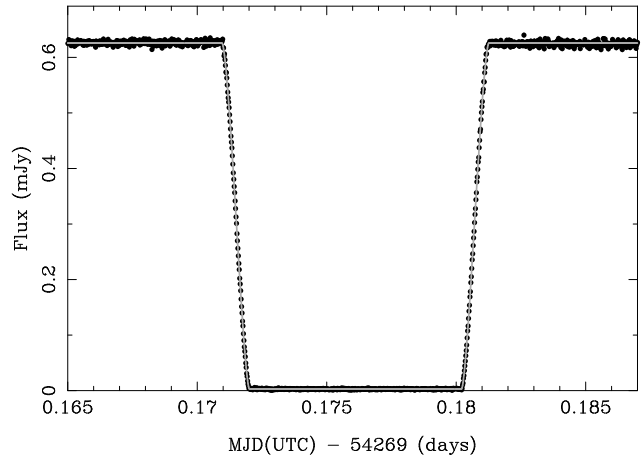}
 \caption{Light curve and model fit to the primary eclipse of GK Vir observed on 2007 June 18.}
 \label{eclfit_1}
\end{center}
\end{figure}

SDSS $0303+0054$ was observed with ULTRACAM in 2007 October, the first run targeted the primary eclipse, the next covered the full orbital period and also covered a flare just after the primary eclipse. The final run again targeted the primary eclipse (though this final run was compromised by poor conditions). The primary eclipses in the \emph{u'},\emph{g'} and \emph{i'} bands are shown in Figure~\ref{priecl2}. The full orbit light curves in the same bands are shown in Figure~\ref{sd0303_full}. Ellipsoidal modulation is visible in the \emph{i'} band, and the \emph{u'} and \emph{g'} band light curves show an increase in the flux up to the primary eclipse then a decrease after the eclipse. This is the opposite to what we would expect if there was a reflection effect and implies that the back side of the secondary star is brighter than the side facing the white dwarf. This may be due to the distribution of spots on the secondary star's surface. 

\subsection{Flaring Rates}
\label{flaresec}
Since ULTRACAM acquires simultaneous images in three different bands, the identification of flares is straightforward. This is due to the fact that flares generally appear largest in the \emph{u'} band and become smaller in redder bands. Table~\ref{flare} lists the number of flares detected for each system throughout all ULTRACAM observations and the total number of hours each system has been observed for. The range of flaring rates given in Table~\ref{flare} is the $90\%$ confidence range ($5\%$ chance of it being lower or higher, based on Poisson statistics). The ULTRACAM data hint that the flaring rates appear to depend on mass rather than rotation rates. The uncertainty in the flaring rates is a result of the small number of hours that these systems have been observed for.

There are several selection effects to consider: flares are easier to see if the white dwarf is cool (faint) and it may be that more massive stars produce brighter flares and so are more visible. It is also possible that flares have been missed, particularly in the fainter systems, if the light curves are particularly noisy. Longer term monitoring of these and similar systems should determine the parameters that dictate flare rates. 

\begin{figure}
\begin{center}
 \includegraphics[width=0.99\columnwidth]{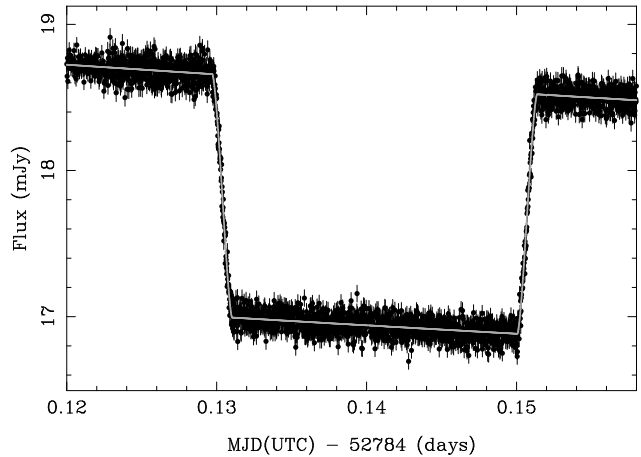}
 \caption{Light curve and model fit to the primary eclipse of RX J$2130.6+4710$ observed on 2003 May 25. A linear slope was added to account for the varying brightness of the secondary star.}
 \label{eclfit_2}
\end{center}
\end{figure}

\begin{table}
 \centering
  \caption{Flaring rates ($90\%$ confidence) for our targets during ULTRACAM observations.}
  \label{flare}
  \begin{tabular}{@{}lccccc@{}}
  \hline
System    & Flares   & Hours  & Flaring & Period & M$_\mathrm{sec}$ \\
          & detected & obs    & rate (hr$^{-1}$)  & (days) & $M_{\sun}$ \\
\hline
RX J2130  & 2        & 8.0    & $0.05-0.38$ & 0.5210 & 0.56  \\
DE CVn    & 2        & 10.5   & $0.04-0.29$ & 0.3641 & 0.41  \\
GK Vir    & 0        & 15.0   & $0.00-0.20$ & 0.3443 & 0.10  \\
SDSS 0110 & 1        & 2.0    & $0.03-1.00$ & 0.3327 & 0.32  \\
RR Cae    & 2        & 3.0    & $0.13-1.00$ & 0.3037 & 0.18  \\
QS Vir    & 2        & 9.0    & $0.04-0.33$ & 0.1508 & 0.43  \\
SDSS 0303 & 1        & 6.0    & $0.01-0.33$ & 0.1344 & 0.25  \\
NN Ser    & 0        & 37.0   & $0.00-0.08$ & 0.1301 & 0.11  \\
\hline
\end{tabular}
\end{table}

\section{O-C Diagrams}
\label{OCtimes}

We wish to determine mid-eclipse times and scaled secondary star radii ($R_\mathrm{sec}/a$) for every recorded ULTRACAM eclipse. We do this by fitting a binary star model. We are not interested in absolute radii but changes in the secondary stars' radii are of interest hence we set the inclination to $90^\circ$ for each system (except NN Ser where the inclination is well known).

For each target we fix the mass ratio ($q$), the inclination ($i$), the white dwarf temperature ($T_\mathrm{WD}$) and the linear limb darkening coefficients for both the white dwarf and the secondary star. The mass ratio and white dwarf temperature are taken from previous studies of each system. The linear limb darkening coefficients are set to 0.2 for each star (except NN Ser for which these are fairly well known, \citealt{parsons10}). By setting the inclination to $90^\circ$, the measured scaled radii are not true radii but represent lower limits on the scaled radius of the secondary star and upper limits on the scaled radius of the white dwarf. This will not affect variations in the radius of the secondary star. 

We initially fitted all the eclipses allowing the two scaled radii, the mid-eclipse time and the temperature of the secondary to vary. The code we used to fit the light curves was designed to produce models for the general case of binaries containing a white dwarf \citep{copperwheat10}. From these fits we determined the mid-eclipse times, we then re-fitted all the eclipses keeping the white dwarf scaled radius fixed at the variance weighted average value from the previous fits. We checked each eclipse fit to ensure a good fit, Figure~\ref{eclfit_1} shows a model fit to an eclipse of GK Vir. For eclipses that have been distorted by the effects of flares we do not determine radii and if the flare significantly affects the eclipse we do not use the mid-eclipse time in our period change analysis. Some of our systems also show a gradient in the light curves across the eclipse as seen in Figure~\ref{eclfit_2} for RX J$2130.6+4710$, due to the varying brightness of the secondary star across its surface, we model this by simply adding a linear slope to our fits.

We find that for most of the systems (with the exception of DE CVn and RR Cae) the scaled radius of the white dwarf determined by fitting the primary eclipses increases as the filter becomes redder. This is due to the fact that we fixed the linear limb darkening coefficient of the white dwarf to 0.2 in all filters. In reality, it is likely that the limb darkening of the white dwarf decreases at longer wavelengths (as was found by \citet{parsons10} for NN Ser), therefore, as seen here, the fits will over-predict the scaled radius of the white dwarf at longer wavelengths. It is interesting to note that the two systems that do not show this trend (DE CVn and RR Cae) contain very cool white dwarfs. Thus, for cool white dwarfs, changes in wavelength apparently do not affect the linear limb darkening coefficients as much as in hotter white dwarfs. 

We correct all our mid-eclipse times to Barycentric Dynamical Time (TDB) and correct for light travel time to the solar system barycentre, thus we use barycentric corrected TDB (BTDB) and list our times in MJD(BTDB). We also convert all previous eclipse times for all our systems to MJD(BTDB) but also list the times in MJD(UTC) and the location of each measurement making corrections to Heliocentric Julian Date (HJD), if required, straightforward. Tables containing all our ULTRACAM eclipse times and secondary star scaled radii, as well as previous eclipse times can be found in the appendix.

\begin{figure}
\begin{center}
 \includegraphics[width=0.99\columnwidth]{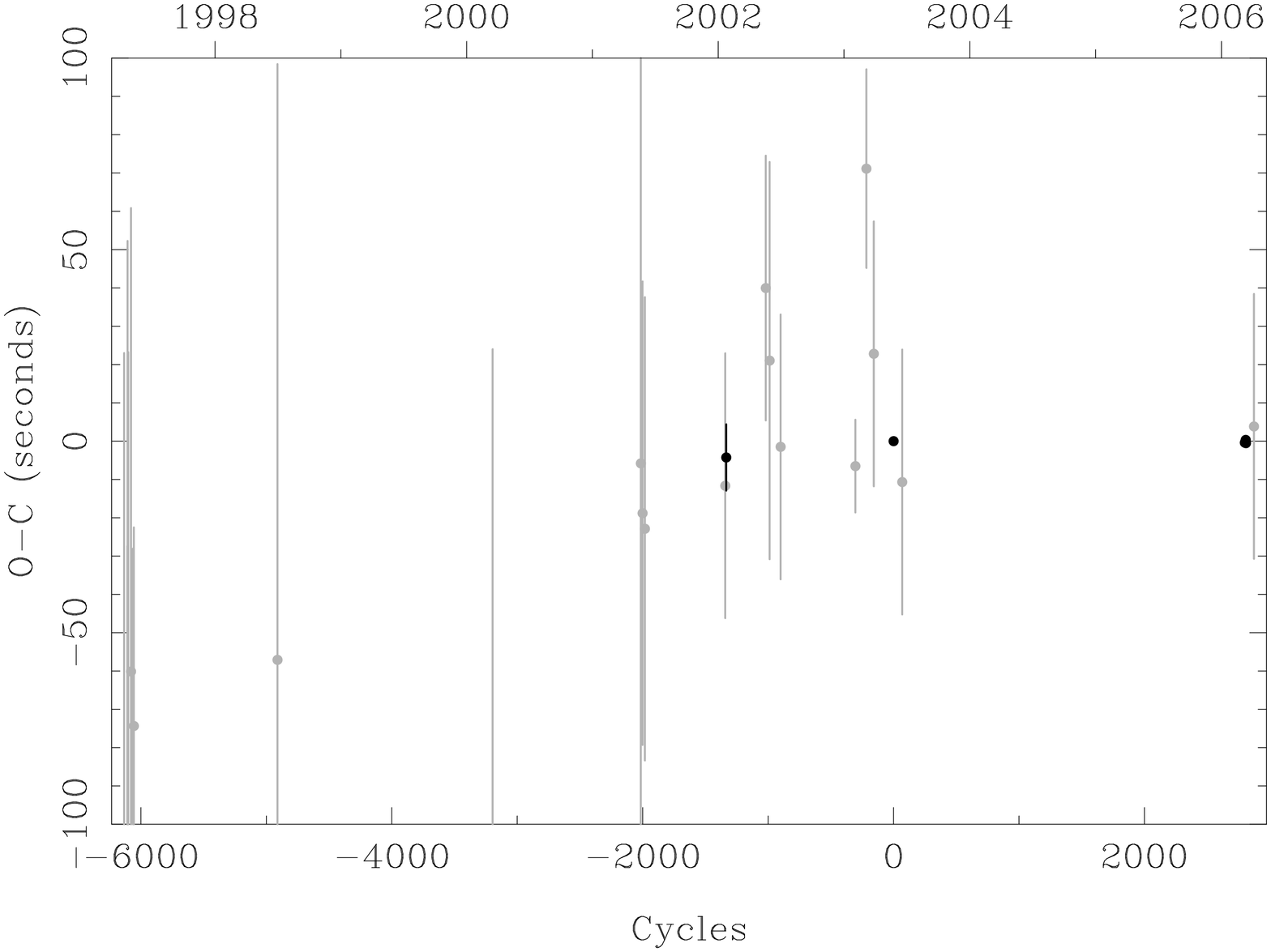}
 \caption{O-C diagram for DE CVn with an ephemeris determined from the ULTRACAM points. Previous data are plotted as open circles whilst the ULTRACAM data are plotted as filled circles, their uncertainties are too small to see at this scale. Measurements with larger errors have been faded. The top axis indicates the year of observation.}
 \label{decvn_oc}
\end{center}
\end{figure}

\subsubsection*{DE CVn}

For fitting the primary eclipses in DE CVn we use a mass ratio of $q=0.80$ and a white dwarf temperature of $T_\mathrm{WD}=8000$ K taken from \citet{besselaar07}. Our fits give average white dwarf scaled radii of $R_\mathrm{WD}/a(\emph{u'})=0.00674(4)$, $R_\mathrm{WD}/a(\emph{g'})=0.00682(3)$, $R_\mathrm{WD}/a(\emph{r'})=0.00732(9)$ and $R_\mathrm{WD}/a(\emph{i'})=0.0069(3)$. where the number in parentheses is the $1\sigma$ error on the last digit.

Table~\ref{decvn_new} lists the mid-eclipse times (in both UTC and BTDB) and the measured secondary star radius for each filter for each eclipse. We also list older eclipse times for DE CVn in Table~\ref{decvn_old} which we have barycentrically corrected (we also list the MJD in UTC). Using just the ULTRACAM points we determine the ephemeris for DE CVn as
\[\mathrm{MJD(BTDB)}= 52784.054\,043(1) + 0.364\,139\,3156(5) E,\]
where the numbers in parentheses are the statistical errors on the last digits. This ephemeris is suitable for predicting future eclipse times though stochastic variations make it likely that these errors will under-predict the true variation in eclipse times.

Figure~\ref{decvn_oc} shows the O-C plot for the eclipse times of DE CVn we have faded those points with larger errors so that any period change is more obvious. Since there are only a few reliable points in the O-C plot we cannot analyse any long term period changes.

\subsubsection*{GK Vir}

For GK Vir we set the mass ratio to $q=0.20$ and the white dwarf temperature to $T_\mathrm{WD}=49000$ K \citep{fulbright93}. Our fits give average white dwarf scaled radii of $R_\mathrm{WD}/a(\emph{u'})=0.00939(3)$, $R_\mathrm{WD}/a(\emph{g'})=0.00949(1)$, $R_\mathrm{WD}/a(\emph{r'})=0.00955(5)$ and $R_\mathrm{WD}/a(\emph{i'})=0.00961(3)$.

\begin{figure}
\begin{center}
 \includegraphics[width=0.99\columnwidth]{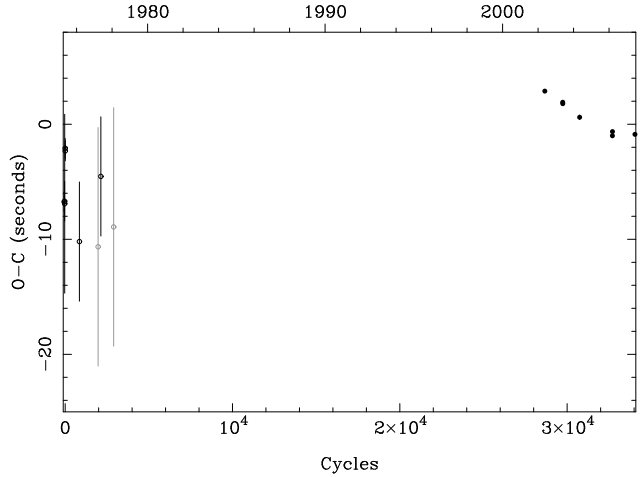}
 \caption{O-C diagram for GK Vir based on our newly derived ephemeris. Previous data are plotted as open circles whilst the ULTRACAM data are plotted as filled circles. Measurements with larger errors have been faded.}
 \label{gkvir_oc}
\end{center}
\end{figure}

Table~\ref{gkvir_new} lists the mid-eclipse times and the measured secondary star radius for each filter and for each eclipse. We also list older eclipse times for GK Vir in Table~\ref{gkvir_old} from \citet{green78}, they corrected their times to Barycentric Julian Date but we believe that light travel time to the solar system barycentre was not taken into account (as applying this $\sim 480$ second correction brings them more into line with our new eclipse times). Our new data help improve the ephemeris of GK Vir, the updated ephemeris is
\begin{eqnarray}
\mathrm{MJD(BTDB)} & = & 42543.337\,7143(30) \nonumber \\
                   &   & +\, 0.344\,330\,838\,759(92) E, \nonumber
\end{eqnarray}
where the numbers in parentheses are the statistical errors on the last two digits. This ephemeris is suitable for predicting future eclipse times but, like DE CVn, larger scale variations may well mean that these errors will under-predict the true variation in eclipse times. The O-C plot for GK Vir is shown in Figure~\ref{gkvir_oc}. Eclipse timings for GK Vir are sparse with none available between 1978 and 2002. The new ULTRACAM points show a period increase between 2002 and 2007, and there has also been a slight variation in O-C times since the earlier observations of \citet{green78}.

\subsubsection*{NN Ser}

In the case of NN Ser we can use precise system parameters hence, we use a mass ratio of $q=0.207$, and an inclination of $i=89.6^{\circ}$, a white dwarf temperature of $T_\mathrm{WD}=57000$ K and linear limb darkening coefficients of 0.125, 0.096, 0.074 and 0.060 for the white dwarf in the \emph{u'}, \emph{g'}, \emph{r'} and \emph{i'} bands respectively and -1.44, -0.48, -0.26 and -0.06 for the secondary star in the same bands \citep{parsons10}. From our initial fits we determine the white dwarf scaled radius as $R_\mathrm{WD}/a(\emph{u'})=0.02262(14)$, $R_\mathrm{WD}/a(\emph{g'})=0.02264(2)$, $R_\mathrm{WD}/a(\emph{r'})=0.02271(10)$ and $R_\mathrm{WD}/a(\emph{i'})=0.02257(10)$.

Table~\ref{nnser_new} lists the mid-eclipse times and the measured secondary star radius for each filter and for each eclipse. Many of these eclipses are the same as in \citet{brinkworth06}; our calculated eclipse times for these eclipses are consistent with theirs. We also list other eclipse times for NN Ser in Table~\ref{nnser_old}, this table includes previous eclipse times and one additional eclipse time taken by us but using ULTRASPEC in imaging mode rather than ULTRACAM. We use the linear ephemeris of \citet{brinkworth06}
\begin{eqnarray}
\mathrm{MJD(BTDB)} & = & 47344.024\,6049(14) \nonumber \\
                   &   & +\, 0.130\,080\,144\,430(36) E \nonumber
\end{eqnarray}
to determine O-C times.

\subsubsection*{QS Vir}

For QS Vir we use a mass ratio of $q=0.66$ and a white dwarf temperature of $T_\mathrm{WD}=14000$ K from \citet{odonoghue03}, for the \emph{r'} and \emph{i'} band eclipses we also fit a slope. From our initial fits we find white dwarf scaled radii of $R_\mathrm{WD}/a(\emph{u'})=0.01297(7)$, $R_\mathrm{WD}/a(\emph{g'})=0.01322(5)$, $R_\mathrm{WD}/a(\emph{r'})=0.01370(12)$ and $R_\mathrm{WD}/a(\emph{i'})=0.01502(37)$.

Table~\ref{qsvir_new} lists the eclipse times for QS Vir from the ULTRACAM data, we also list all previous eclipse times in Table~\ref{qsvir_old}. We determine the eclipse times from \citet{odonoghue03} by averaging their mid-ingress and mid-egress times then converting to BTDB; we convert the mid-eclipse times of \citet{qian10} from UTC to BTDB. We also list our other eclipse times not observed with ULTRACAM. For eclipse cycles 43342 and 43362 the observations start during the eclipse therefore we determine the mid-eclipse times by measuring the centre of the egress then applying a correction based on our eclipse model. A minor earthquake occurred during the egress of eclipse cycle 43349 causing the loss of some data, nevertheless enough data were available to determine a mid-eclipse time. We determine O-C times using the linear ephemeris of \citet{odonoghue03} corrected to BTDB
\[\mathrm{MJD(BTDB)} = 48689.140\,62(1) +\, 0.150\,757\,525(1) E.\]

\subsubsection*{RR Cae}

To fit the two ULTRACAM eclipses of RR Cae we use a mass ratio of $q=0.42$ and a white dwarf temperature of $T_\mathrm{WD}=7540$ K taken from \citet{maxted07}. Our fits give an average white dwarf scaled radius of $R_\mathrm{WD}/a(\emph{u'})=0.01436(18)$, $R_\mathrm{WD}/a(\emph{g'})=0.01448(2)$ and $R_\mathrm{WD}/a(\emph{i'})=0.01433(9)$.

The new ULTRACAM eclipse times and measured secondary star radii are listed in Table~\ref{rrcae_new} and previous eclipse times are shown in Table~\ref{rrcae_old}. We use the ephemeris of \citet{maxted07}
\[\mathrm{MJD(BTDB)} = 51522.548\,5670(19) +\, 0.303\,703\,6366(47) E,\]
to calculate the O-C times for RR Cae. 

\subsubsection*{RX J2130.6+4710}

For RX J$2130.6+4710$ we use the parameters of \citet{maxted04} namely, $q=1.00$ and $T_\mathrm{WD}=18000$ K. We include a slope in the \emph{i'} band eclipse fits. Our initial fits give white dwarf scaled radii of $R_\mathrm{WD}/a(\emph{u'})=0.00768(3)$, $R_\mathrm{WD}/a(\emph{g'})=0.00775(2)$ and $R_\mathrm{WD}/a(\emph{i'})=0.00785(8)$.

The ULTRACAM eclipse times and secondary star radii measurements of RX J$2130.6+4710$ are shown in Table~\ref{rxj_new} and other eclipse times in Table~\ref{rxj_old}. The eclipse observed in 2002 (cycle -716) featured a flare during the egress, hence we do not determine the secondary star's radius for this eclipse. Our ULTRACAM eclipses are the same as in \citet{maxted04} and are consistent with their results, though we apply a light travel time correction to their times to put them in MJD(BTDB). We also re-iterate the warning made in \citet{maxted04} that all the eclipse times around cycle -1900 may be in error by a few seconds and should not be used to study any long-term period changes. 

We use the ephemeris of \citet{maxted04} and correct it to MJD(BTDB)
\[\mathrm{MJD(BTDB)} = 52785.182\,620(2) +\, 0.521\,035\,625(3) E.\]
The O-C times for RX J$2130.6+4710$ give an identical plot to that shown in \citet{maxted04} since no additional eclipse times are available. Little can be taken from the O-C times since the current number of eclipse times is still quite sparse, hence additional eclipse times are required before any detailed analysis of the period changes in RX J$2130.6+4710$ can be made.

\subsubsection*{SDSS 0110+1326}

We use a mass ratio of $q=0.54$ and a white dwarf temperature of $T_\mathrm{WD}=25900$ K \citep{pyrzas09} to fit the single observed eclipse of SDSS $0110+1326$. Since there is only one ULTRACAM eclipse we only fit it once determining white dwarf scaled radii of $R_\mathrm{WD}/a(\emph{u'})=0.01415(16)$, $R_\mathrm{WD}/a(\emph{g'})=0.01431(4)$ and $R_\mathrm{WD}/a(\emph{i'})=0.01426(15)$. 

Table~\ref{sdss0110_new} details the ULTRACAM eclipse. We also list previous eclipse times in Table~\ref{sdss0110_old}. We calculate the O-C times using the ephemeris of \citet{pyrzas09} corrected to MJD(BTDB)
\[\mathrm{MJD(BTDB)} = 53993.948\,65(9) +\, 0.332\,687\,3(1) E.\]
The ULTRACAM eclipse time is the most accurate published to-date for this system and shows some deviation from the ephemeris, however, given the large errors on those points used to determine the ephemeris, further accurate eclipse times are likely to greatly improve the ephemeris for SDSS $0110+1326$. Since there is only one precise eclipse time the analysis of any period changes in this system will have to wait until further data are available.

\subsubsection*{SDSS 0303+0054}

For SDSS $0303+0054$ we adopt a mass ratio of $q=0.28$ and a white dwarf temperature of $T_\mathrm{WD}=8000$ K \citep{pyrzas09}. Our initial fits give white dwarf scaled radii of $R_\mathrm{WD}/a(\emph{u'})=0.0093(7)$, $R_\mathrm{WD}/a(\emph{g'})=0.0098(1)$ and $R_\mathrm{WD}/a(\emph{i'})=0.0100(5)$. 

The new ULTRACAM eclipse times and measured secondary star scaled radii are listed in Table~\ref{sdss0303_new}, poor conditions led to the loss of data in the \emph{u'} band during eclipse cycle 3058. We also list all previous eclipse times for SDSS $0303+0054$ in Table~\ref{sdss0303_old} and use the ephemeris of \citet{pyrzas09} corrected to MJD(BTDB)
\[\mathrm{MJD(BTDB)} = 53991.117\,18(10) +\, 0.134\,437\,72(7) E.\]
to determine the O-C times. Our three new eclipse times are the most accurate for this system so far and, given the large uncertainty in the ephemeris, agree well with previous eclipse times. However, the small number of precise eclipse times means that any long term period changes are not yet visible in the data.

\section{Discussion}

\subsection{The period change of QS Vir}

\begin{figure*}
\begin{center}
 \includegraphics[width=0.9\textwidth]{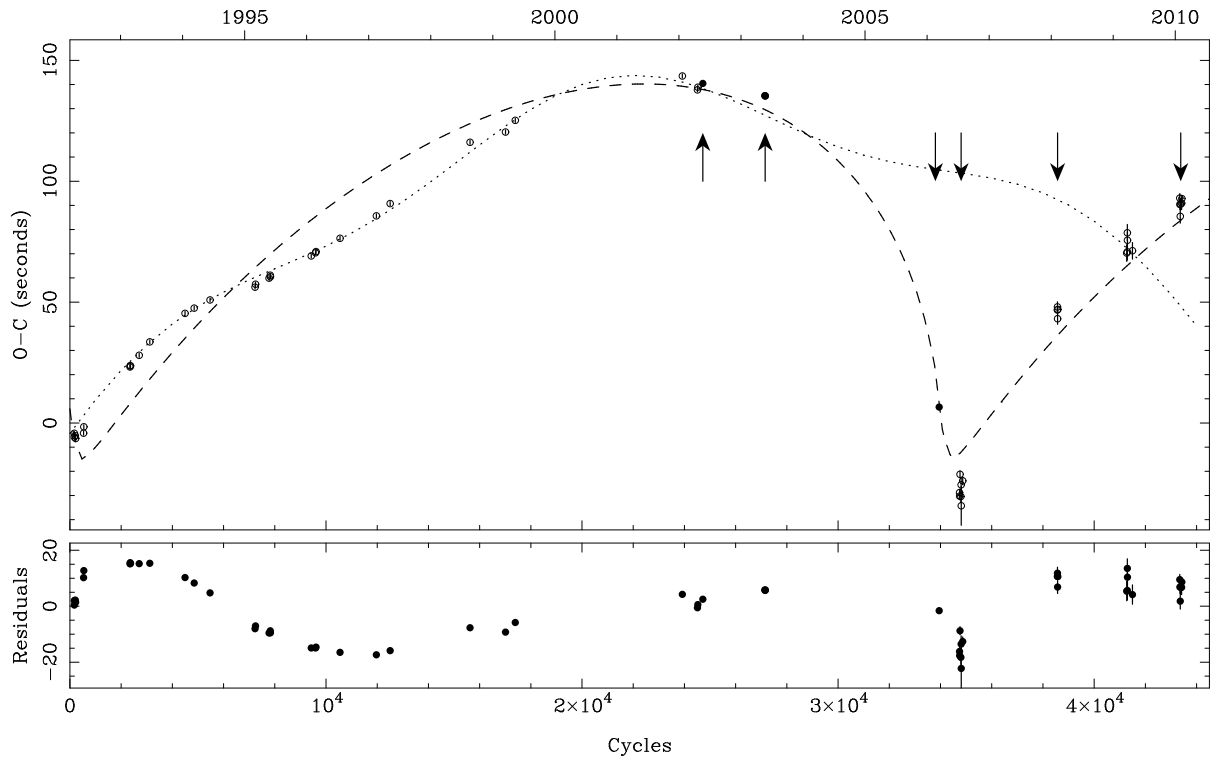}
 \caption{\emph{Top:} O-C diagram for QS Vir. The ephemeris was determined by varying the ephemeris of \citet{odonoghue03} in order to fit a third body in an elliptical orbit (\emph{dashed} line). We also include the sinusoidal fit from \citet{qian10} (\emph{dotted} line) . Our additional eclipse times, indicated by the arrows, clearly disagree with the sinusoidal fit. \emph{Bottom:} residuals of the fit to the third body in an elliptical orbit. The filled circles are the ULTRACAM eclipse points.}
 \label{qsvir_oc}
\end{center}
\end{figure*}

The O-C plot for QS Vir is shown in Figure~\ref{qsvir_oc}, the eclipse times show a substantial shift after $\sim 20,000$ cycles. \citet{qian10} used their new eclipse times together with previous times and fitted a sinusoid to them. This fit is the dotted line in Figure~\ref{qsvir_oc}. Clearly our new eclipse times, indicated by arrows, disagree strongly with this fit. Hence we conclude that, as with NN Ser, the proposed planet in QS Vir does not exist. 

The eclipse times show a complex behaviour, and to see if this period change could be caused by Applegate's mechanism we measured the maximum period shift as $\sim$0.05 seconds in $\sim$2 years (in the region where the O-C times turn, around cycle number 35,000). We use Applegate's equation for the energy required to generate a period change
\begin{eqnarray}
\label{apple1}
\Delta E = \Omega_\mathrm{dr} \Delta J + \frac{\Delta J^{2}}{2 I_\mathrm{eff}},
\end{eqnarray}
where $\Omega_\mathrm{dr}$ is the initial differential rotation which we set to zero since we are after the minimum energy required to produce this period change. The star is separated into a shell and a core, $I_\mathrm{eff} = I_{S}I_{*}/(I_{S}+I_{*})$ is the effective moment of inertia where $S$ stands for the shell and $*$ represents the core. We follow the prescription of \citet{applegate92} and set the shell mass to $M_{S} = 0.1M_{\sun}$ meaning that $I_\mathrm{eff} = 0.5 I_{S} = (1/3) M_{S} R_{*}^{2}$. The change in angular momentum, $\Delta J$, is calculated via
\begin{eqnarray}
\label{apple2}
\Delta J = \frac{-G M^{2}}{R} \left( \frac{a}{R} \right)^{2} \frac{\Delta P}{6 \pi}
\end{eqnarray}
using the mass and radius of the secondary star and orbital separation from \citet{odonoghue03} namely $M = 0.51 M_{\sun}$, $R = 0.42 R_{\sun}$ and $a = 1.28 R_{\sun}$. We determine that the minimum energy required to drive the maximum observed period change in QS Vir is $3.0 \times 10^{40}$ ergs. The  luminosity of the secondary star is given by $L = 4 \pi R^{2} \sigma T^{4}$ which over the two years supplies $3.5 \times 10^{39}$ ergs, failing by an order of magnitude to explain the observed period change. This is likely to be even worse if we apply the generalised version of Applegate's calculation introduced in \citet{brinkworth06}.

Another explanation for the observed shift in eclipse times is a third body. If the third body is in a highly elliptical orbit then for much of its orbit the eclipse times will remain roughly constant but as the third body swings inward the central binary moves towards and away from us quickly resulting in a large, short-lived timing change. 

We fit an elliptical orbit to the eclipse times, allowing the ephemeris of QS Vir to change, and determine that a third body with a minimum mass of $M\sin{i} \sim 0.05 M_{\sun}$ in a $\sim 14$ year orbit with an eccentricity of $\sim 0.9$ approximately fits the data, this fit is the dashed line in Figure~\ref{qsvir_oc} for an inclination of $90^{\circ}$. The linear ephemeris obtained from this fit is
\[\mathrm{MJD(BTDB)} = 48689.141\,163(10) +\, 0.150\,757\,453(1) E.\]

Since this system has undergone substantial evolution the existence of a third body in such an orbit is questionable. To see if such an orbit is possible, we must analyse the history of QS Vir. We estimate the minimum progenitor mass of the white dwarf to be $\sim 1.8 M_{\sun}$ \citep{meng08}, with a core mass equal to the current white dwarf mass ($0.77 M_{\sun}$). This corresponds to a radius on the asymptotic giant branch (AGB) of $\sim 460 R_{\sun}$ \citep{hurley00}. We can calculate the initial separation of the binary from the \citet{eggleton83} formula
\begin{eqnarray}
\label{eggle}
R_{L} = \frac{0.49 q^{2/3} a_{i}}{0.6 q^{2/3} + \ln(1 + q^{1/3})}
\end{eqnarray}
and setting $R_{L} = R_\mathrm{AGB}$, where $q = M_\mathrm{AGB}/M_{2}$ and $M_{2}$ is the mass of the secondary star. This gives an initial binary separation of $a_{i} = 4.4$ AU. The fit to the eclipse times implies the current semimajor axis of the third body is $\sim 6.4$ AU, assuming an adiabatic change in semimajor axis during the mass loss phase of the primary, implies that the semimajor axis of the third body before the common envelope phase was $\sim 3.6$ AU. By altering the period of QS Vir a longer period fit can be obtained but it requires a similarly high eccentricity and is a slightly poorer fit, and still results in a very small periapsis separation. Since the eccentricity of the third body should have been little affected by the mass loss \citep{jeans24} all these possible orbits cross the orbit of the secondary star meaning that it is unlikely to have survived for the entire main sequence lifetime of the primary. In addition, since the common envelope must have reached out to at least the secondary star, the orbit of this third body would have taken it into the common envelope resulting in a dramatically different orbit to what we now see. 

It also appears doubtful that the third body formed out of the material in the common envelope. A similar mechanism has been used to explain the creation of planets around pulsars \citep{lin91} out of the supernova material. However, the high eccentricity and mass of this object would seem to make creation via such a mechanism unlikely. However, since the dynamics of the system throughout the common envelope phase are subject to large uncertainties, we cannot rule out the existence of this third body and it remains the only mechanism able to produce such a large period variation.

The residuals of the elliptical orbit fit, shown in the bottom panel of Figure~\ref{qsvir_oc}, still show considerable structure, but they are at a level consistent with Applegate's mechanism. Further monitoring of the eclipse times may reveal the true nature of this remarkable period change.

\subsection{The period change of NN Ser}

\begin{figure*}
\begin{center}
 \includegraphics[width=0.90\textwidth]{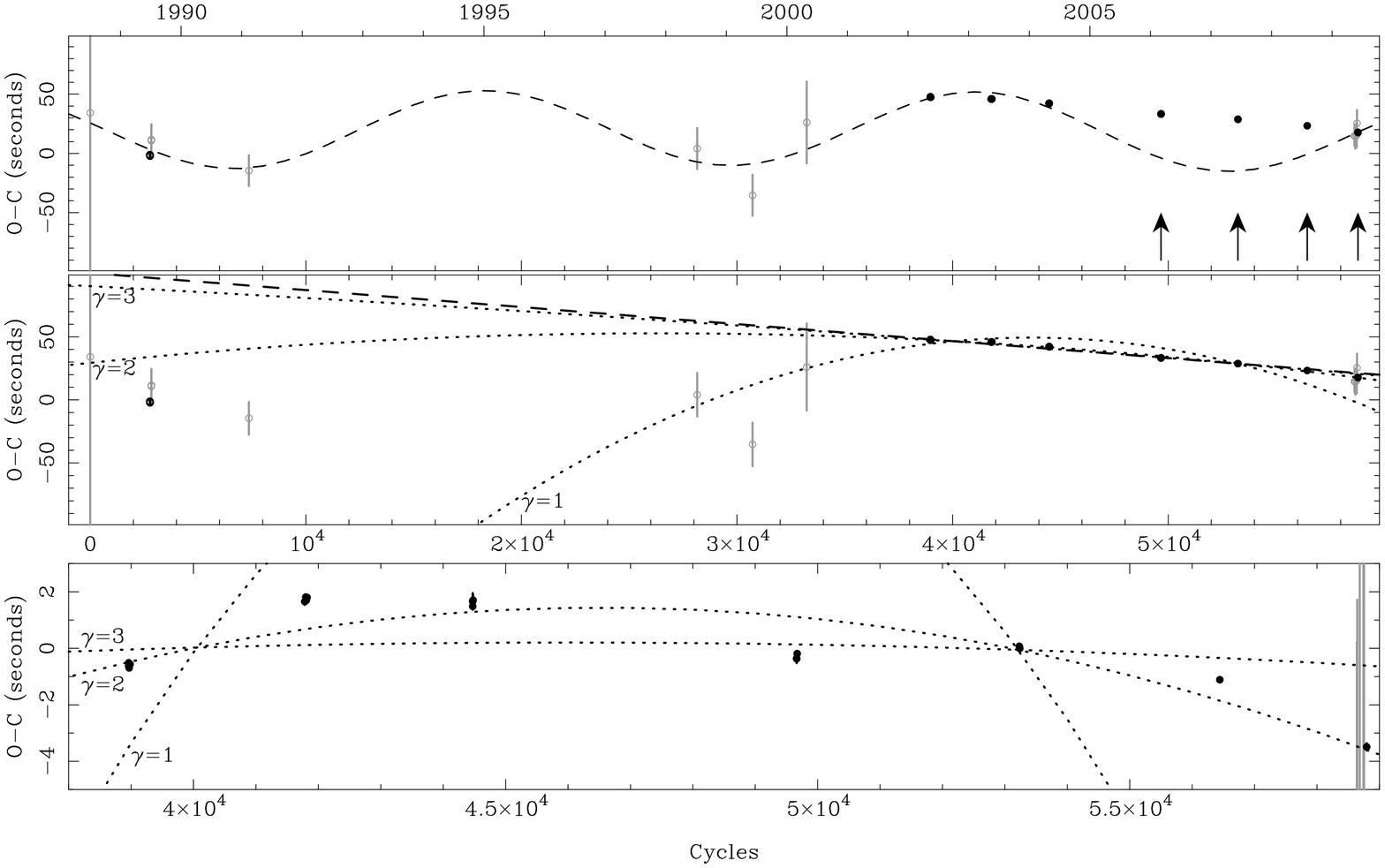}
 \caption{\emph{Top:} O-C diagram for NN Ser based on the ephemeris of \citet{brinkworth06}, with sinusoidal fit from \citet{qian09} over plotted. The additional ULTRACAM points, indicated by the arrows, clearly disagree with the fit (the last point coincides with the points from \citet{qian09} but has a error comparable in size to the other ULTRACAM points). \emph{Centre:} O-C diagram with a linear fit to just the ULTRACAM points (\emph{dashed} line) and standard magnetic braking models (\emph{dotted} lines). \emph{Bottom:} residuals of the linear fit to the ULTRACAM points with the same magnetic braking models (\emph{dotted} lines). Previous data are plotted as open circles whilst the ULTRACAM data are plotted as filled circles. Measurements with larger errors have been faded.}
 \label{nnser_oc}
\end{center}
\end{figure*}

\citet{qian09} proposed the existence of a planet in NN Ser based on eclipse timings. The top panel of Figure~\ref{nnser_oc} shows their sinusoidal fit along with all eclipse times. Our new times, which we indicate with arrows, clearly disagree with the sinusoidal fit, hence we conclude that the third body proposed by \citet{qian09} doesn't exist. We fit a linear ephemeris to just the ULTRACAM points, the centre panel of Figure~\ref{nnser_oc} shows this fit (the \emph{dashed} line). We determine 
\begin{eqnarray}
\mathrm{MJD(BTDB)_{UCAM}} & = & 47344.025\,768\,43(96) \nonumber \\
                   &   & +\, 0.130\,080\,115\,390(20) E \nonumber
\end{eqnarray}
from the ULTRACAM points. The bottom panel of Figure~\ref{nnser_oc} shows the residuals of this fit around the ULTRACAM points. Additional small scale variations are visible in this plot which are most likely the result of Applegate's mechanism.

The period change of NN Ser was analysed by \citet{brinkworth06}, who determined that Applegate's mechanism fails to explain the large period change. They determined that if magnetic braking is not cut off below $0.3M_{\sun}$ then it can explain the period change. We use the standard magnetic braking relationship from \citet{rappaport83}
\begin{eqnarray}
\label{magbrak}
\dot{J} \approx -3.8 \times 10^{-30} M_{\sun} R_{\sun}^{4} m_{2} r_{2}^{\gamma} {\omega}^{3} \, \mathrm{erg},
\end{eqnarray}
where $m_{2}$ and $r_{2}$ are the secondary star's mass and radius and $\omega$ is the angular frequency of rotation of the secondary star. $\gamma$ is a dimensionless parameter which can have a value between 0 and 4. We determine the angular momentum loss using Equation~\ref{magbrak} and the parameters from \citet{parsons10}, then use this to fit a parabola to the ULTRACAM data points. We use a range of values for $\gamma$; for $\gamma=4$ we find that the period change is negligible whilst for $\gamma=0$ the period change is far higher than observed. We also calculated the period change using the relationship given by \citet{verbunt81} and for gravitational radiation \citep{peters64} but find that both these methods give a negligible period change. In the context of cataclysmic variable evolution $\gamma=2$ is frequently used \citep{schreiber03}. For this value we find a good fit to the ULTRACAM points, however, this fit passes somewhat above earlier points suggesting $\gamma \sim 1.8$ if this is indeed the explanation. It should be stressed that these relationships are by no means proven but we show them here as a possible explanation for the observed period change. 

The analysis of our new eclipse times agrees with the conclusions made by \citet{brinkworth06} that the only mechanisms able to explain the observed period change in NN Ser are magnetic braking (provided it is not cut off below $0.3M_{\sun}$) or perhaps the existence of a third body in a long period orbit around NN Ser.

\subsection{The period change of RR Cae}

\begin{figure}
\begin{center}
 \includegraphics[width=0.99\columnwidth]{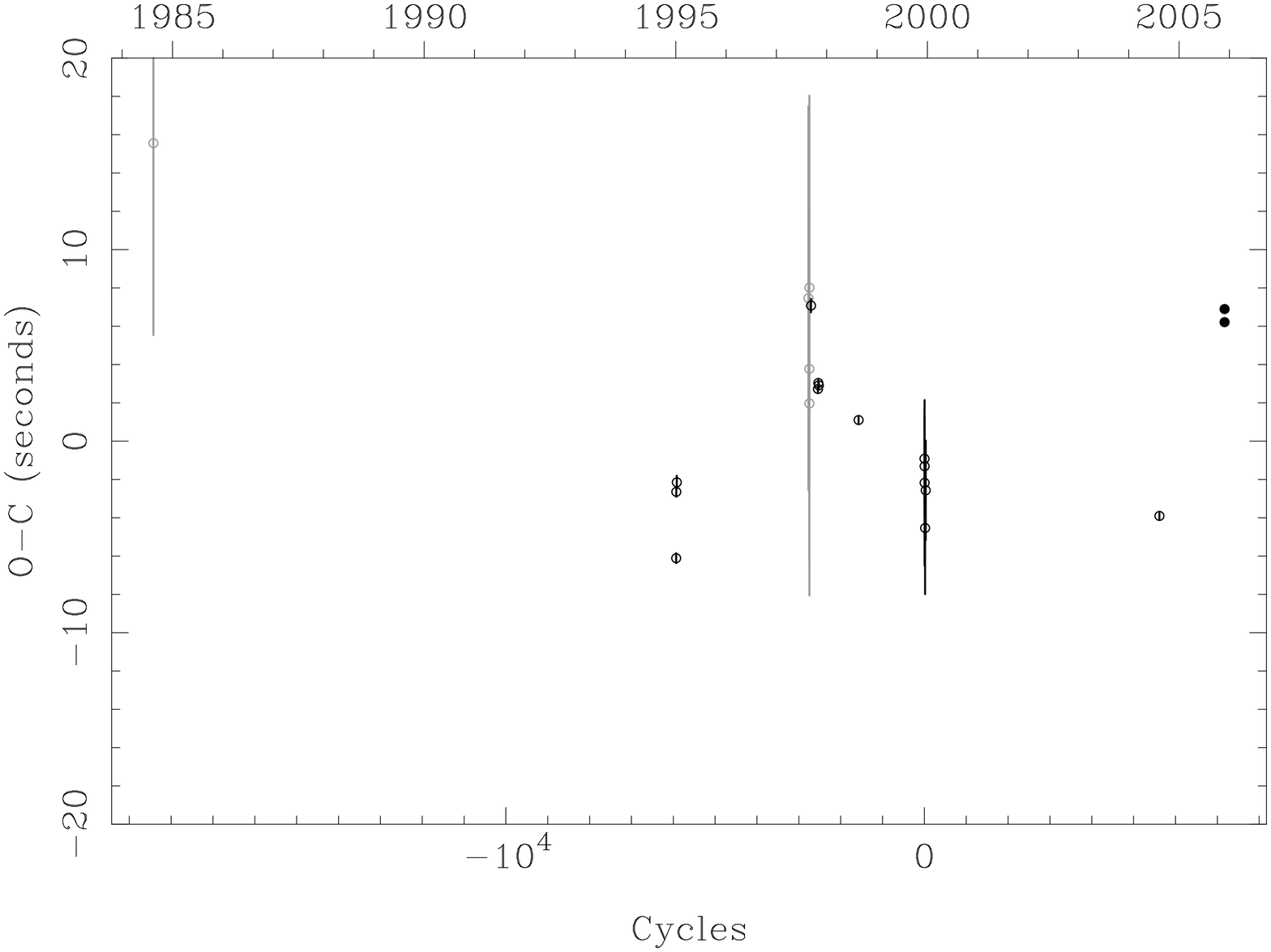}
 \caption{O-C diagram for RR Cae based on the ephemeris of \citet{maxted07}. Previous data are plotted as open circles whilst the ULTRACAM data are plotted as filled circles. Measurements with larger errors have been faded. Eclipse cycle -2760 shows large scatter compared to all the other points hence we do not include it in this figure.}
 \label{rrcae_oc}
\end{center}
\end{figure}

The O-C plot for RR Cae is shown in Figure~\ref{rrcae_oc}. It shows a roughly sinusoidal variation; in order to see if Applegate's mechanism is able to drive these small period changes, we use a similar analysis to that of QS Vir. The change in period from cycles -5900 to 0 ($1.6 \times 10^{8}$ sec) is $\sim 0.006$ sec. We use the system parameters from \citet{maxted07} namely, $M_\mathrm{sec} = 0.182 M_{\sun}$, $R_\mathrm{sec} = 0.215 R_{\sun}$, $a = 1.623 R_{\sun}$ and $T_\mathrm{sec} = 3100$ K. In order to drive the observed period change we require $\sim 3.8 \times 10^{39}$ ergs. Over the observed time period the secondary star produces $\sim 2.4 \times 10^{39}$ ergs which, given the uncertainty in the system parameters and the fact that this is only a rough calculation, demonstrates that Applegate's mechanism is able to explain the observed period change in RR Cae.

It is likely that the eclipse times in RR Cae are being affected by micro-flares that are only visible near the bottom of the primary eclipse. At blue wavelengths these distort the shape of ingress and egress and so produce jitter of up to several seconds in the individual eclipse timings. Further eclipse times may show a descrepancy between the \emph{u'} band eclipse times and the redder band times.

There is little evidence in the eclipse times of long term period change via angular momentum loss. This is unsurprising given that \citet{maxted07} calculated that the period change would be of the order of $5 \times 10^{-14} < \dot{P}/P < 1.4 \times 10^{-13}$ depending upon which magnetic braking prescription is used. However, additional precise eclipse times may reveal this change in the future. 

\subsection{Variations in Secondary Star Radii}

For the systems GK Vir and NN Ser we have accurate measurements of the secondary star's radius spanning five and six years respectively. The other systems require more measurements before any potential trends can be identified. 

A starspot's reduced pressure, density and temperature with respect to its surroundings results in its depression below the surrounding photosphere by several hundreds of kilometres. This effect is known as a Wilson depression. The presence of a Wilson depression on the limb of a secondary star as it occults the primary may cause small changes in the O-C times since it may delay the time of eclipse ingress or advance the time of eclipse egress. \citet{watson04} showed that this effect can cause small jitters in the O-C times of up to a few seconds.

A Wilson depression causes a small decrease in the eclipse duration and also displaces the measured centre of the eclipse, hence we would expect that the duration of the eclipse and the jitter in O-C times would be correlated were there Wilson depressions present. For both NN Ser and GK Vir we find no evidence of such a correlation leading us to conclude that the eclipse times of these two systems are not affected by Wilson depressions. The fact that both of the secondary stars in these systems have never shown any flaring events supports this and shows that both these rapidly rotating stars are remarkably quiet.  

\begin{figure*}
\begin{center}
 \includegraphics[width=0.99\columnwidth]{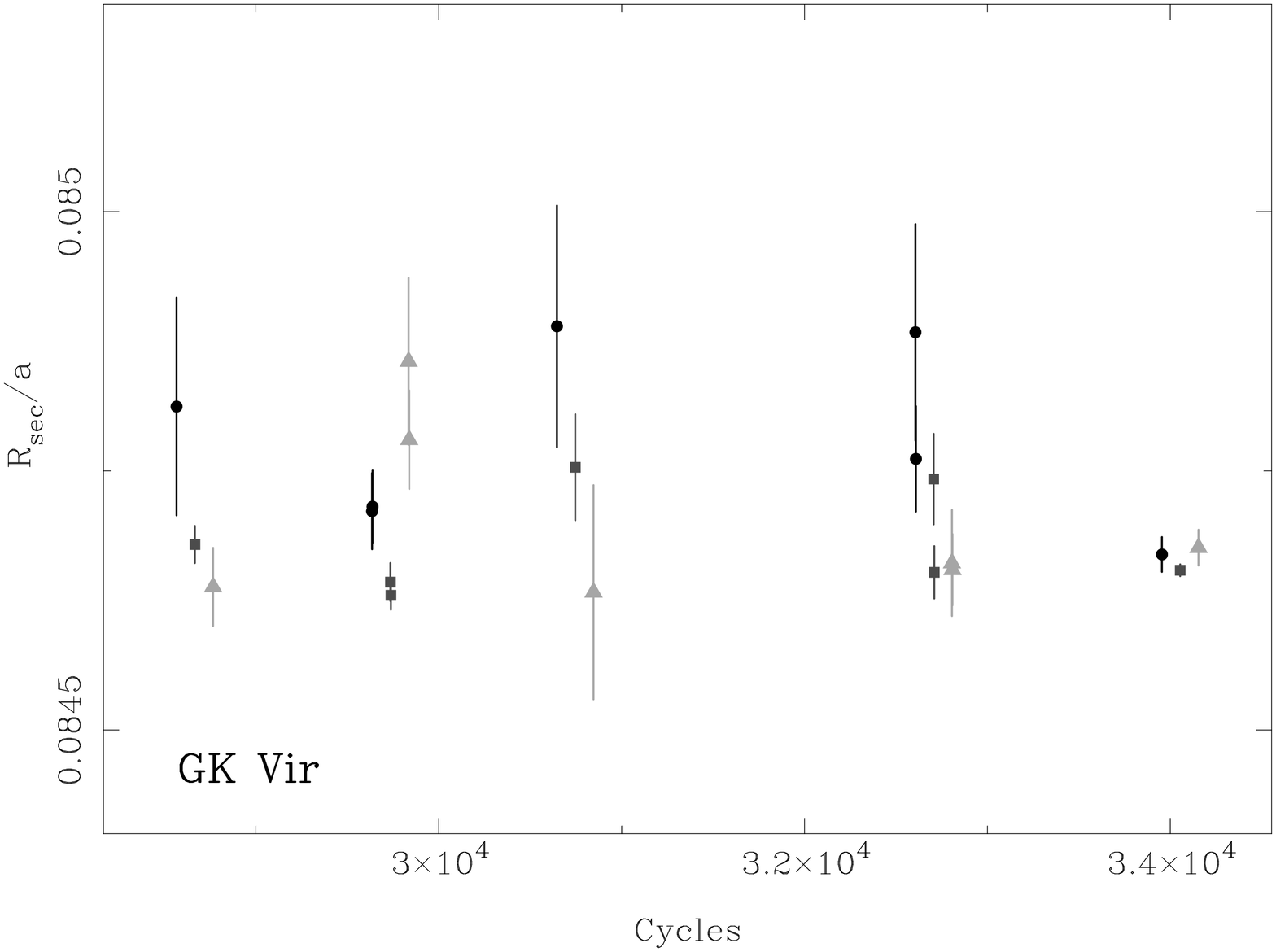}
 \includegraphics[width=0.99\columnwidth]{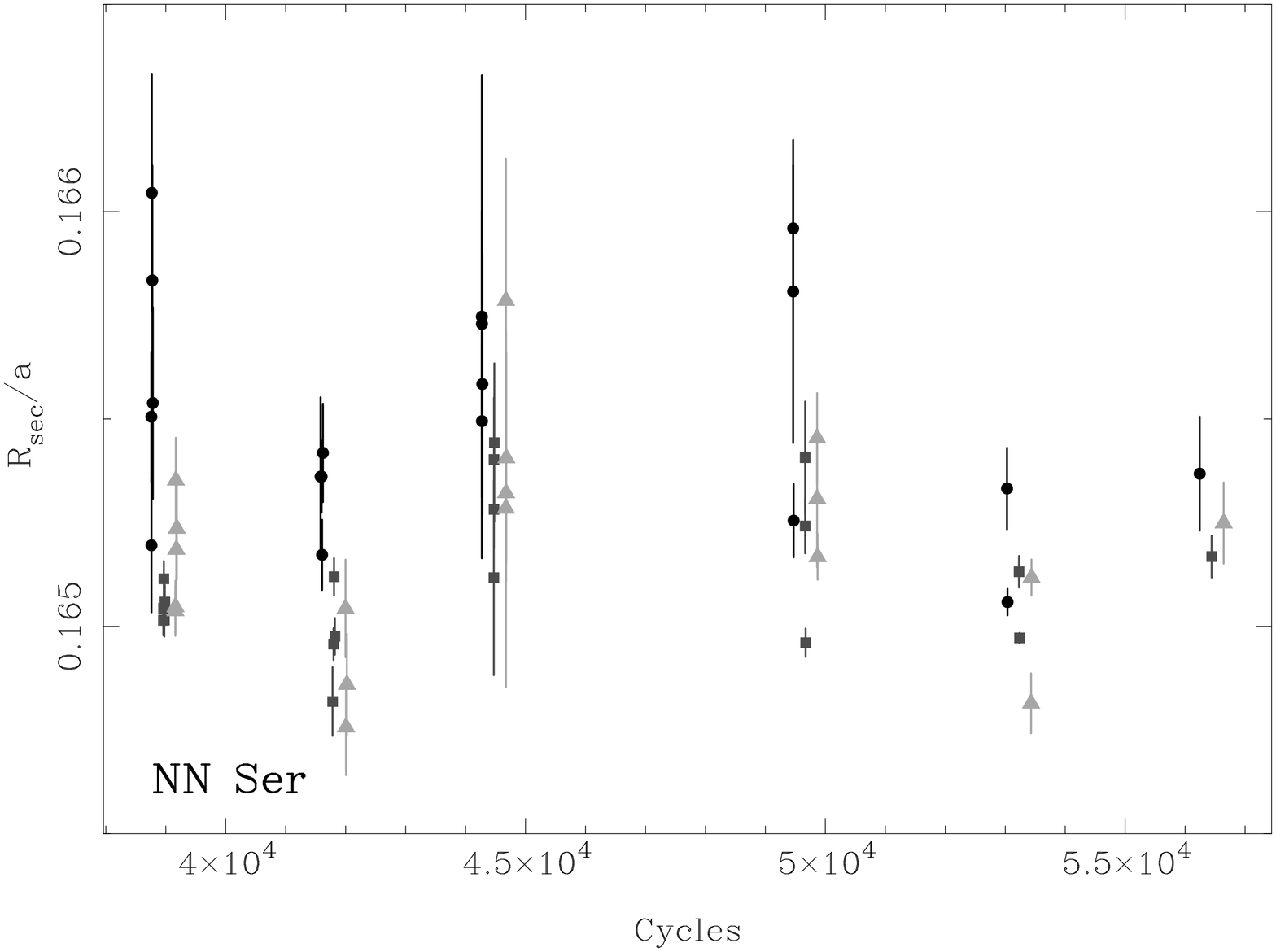}
 \caption{Measured secondary star scaled radii for GK Vir (\emph{left}) and NN Ser (\emph{right}). The \emph{black circles} are the \emph{u'} band measurements and have been offset by -100 cycles for GK Vir and -200 cycles for NN Ser. The \emph{dark grey squares} are the \emph{g'} band measurements and the \emph{light grey triangles} are the red band (\emph{r'} or \emph{i'}) measurements and have been offset by +100 cycles for GK Vir and +200 cycles for NN Ser.}
 \label{secrad}
\end{center}
\end{figure*}

Applegate's mechanism can also effect the duration of the eclipse since the result of this mechanism is to alter the oblateness of the star. \citet{applegate92} calculated that the deformation of the star, $\psi$, via this mechanism is
\begin{eqnarray}
\label{deform}
\frac{\psi}{R_\mathrm{sec}} = \frac{1}{3} \frac{\Omega^{2} R_\mathrm{sec}^{3}}{G M_\mathrm{sec}}
\end{eqnarray}
where $M_\mathrm{sec}$ and $R_\mathrm{sec}$ are the mass and radius of the secondary star and $\Omega$ is its angular velocity. However, since this is the deformation at the sub-stellar point and the poles, inclinations where the primary passes across the face of the secondary between these extremes will result in a smaller observed deformation, hence this represents an upper limit. For NN Ser, using the system parameters of \citet{parsons10} gives a deformation of $\psi \sim 10^{-3} R_\mathrm{sec}$. Using the parameters of \citet{fulbright93} gives a deformation for GK Vir of $\psi \sim 10^{-4} R_\mathrm{sec}$.

Figure~\ref{secrad} shows the variation in secondary star radius for GK Vir and NN Ser over the period of ULTRACAM observations. For GK Vir, the \emph{u'} band measurements have been offset by -100 cycles and the red band measurements (\emph{r'} or \emph{i'}) by +100 cycles. There does not appear to be any variation in the size of the secondary star. However, the 2007 observations made at the VLT are the last set of points ($\sim$ cycle 34000) and are extremely precise with $\Delta R_\mathrm{sec} / R_\mathrm{sec} < 10^{-5}$. Additional points with precisions of this order might be able to detect changes in the radius of the secondary star as a result of Applegate's mechanism.

For NN Ser, the \emph{u'} band measurements have been offset by -200 cycles and the red band measurements (\emph{r'} or \emph{i'}) by +200 cycles. The measured secondary star scaled radius appears to show a very slight variation of order the size we would expect from Applegate's mechanism however, the errors are too large to be sure. Additional measurements with the accuracy of the best points might reveal any underlying variations. 

The accuracy of these measurements is encouraging and potentially offers us a method of independently verifying Applegate's mechanism. These two systems are particularly useful in this regard as the secondary stars in both systems show no signs of activity. For those systems that do show substantial activity, such as DE CVn, QS Vir, RR Cae and RX J$2130.6+4710$, Wilson depressions may affect the eclipses and this may become evident with additional measurements of the width of eclipses. The deformation induced by Applegate's mechanism is also larger for the stars in these systems hence accurate additional monitoring of these systems may identify this effect.

\subsection{Detecting planets in eclipsing compact binaries}

Detection of extrasolar planets via timing observations have been successful around pulsars (see for example \citealt{ford00} and \citealt{konacki03}) five planets have been confirmed with this method. Recently these methods have been used to study the eclipse times of compact binaries such as PCEBs since the O-C times in these systems will be affected by the presence of any third body. The possibility of sub-stellar components in these systems suggests intriguing questions about both their history as well as the history of the system as a whole.

However, we have shown that the sub-stellar components suggested as the cause of the O-C variations in NN Ser and QS Vir are incompatible with our new eclipse times and hence do not exist. It seems that additional eclipse times invariably disagree with previous sinusoidal fits, hence regularly sampled eclipse times are essential. This is particularly striking in the case of QS Vir where the previously poorly sampled eclipse times between 2002 and 2009 lead \citet{qian10} to miss the large deviation from linearity. Similar issues may affect the detection of multiple planets via transit time variations (\citealt{watson10}, in press). 

The stability of any additional companions to PCEBs must be studied over the full history of the system. A simple calculation of the orbital configuration of the system before the common envelope phase started will show that some systems cannot have existed during this phase. For example, following the same procedure used to analyse the potential third body found in QS Vir, we take the parameters for the sub-stellar component in NN Ser proposed by \citet{qian09} and determine that the semi-major axis of the sub-stellar component before the common envelope phase was $\sim 1.6$ AU which was smaller than that of the secondary star ($\sim 1.8$ AU). Therefore the system could not have evolved to its present configuration since these two objects would have had to have crossed each others path. A similar situation is found for the sub-stellar companion thought to exist in QS Vir.

It may be possible to form planets out of the common envelope material thus getting around some of these evolutionary problems, but this mechanism creates additional problems since the created body must still move to its current location.

In light of these findings, we advise caution when using eclipse times of compact binaries to detect planets. Eclipse times must be regularly sampled over long time periods and the history of any third body must be analysed to check its stability. Reliable detections of third-bodies will unfortunately require many decades of monitoring. We also require better understanding of the other processes that can cause period changes. Confirmation of any proposed planetary companions to these systems must come by other methods (radial velocity variations, planetary transits etc.). 

\section{Conclusions}

We have presented high time resolution ULTRACAM light curves for the systems DE CVn, GK Vir, NN Ser, QS Vir, RR Cae, RX J$2130.6+4710$, SDSS $0110+1326$ and SDSS $0303+0054$. By fitting models to all the observed eclipses we were able to determine extremely accurate mid-eclipse times, which we combine with earlier eclipse times to determine any period changes. We found that the conclusions made by \citet{brinkworth06} are still true for NN Ser namely, that Applegate's mechanism fails to explain the observed period change but that magnetic braking can, but given the low mass of the secondary star in NN Ser this requires that magnetic braking is not cut off below $0.3 M_{\sun}$ raising problems for cataclysmic variable evolution if true. Additionally, we determine that small period variations observed in RR Cae can be generated via Applegate's mechanism.

We detect a 250 second departure from linearity in the eclipse times of QS Vir which is best fit by a combination of a third body ($M \sim 0.05 M_{\sun}$) in an eccentric orbit and Applegate's mechanism. A simple analysis of the system's past implies that this potential companion would most likely have interacted with the common envelope making the current system arrangement unlikely, however, given the uncertainties involved in the common envelope stage we cannot rule out the existence of this third body. If confirmed, this third body may offer some insight into common envelope evolution. We also detect smaller period variations which can be explained as the result of Applegate's mechanism. 

Our eclipse times also show that the two sub-stellar components proposed to exist in NN Ser and QS Vir do not exist. We conclude that great care must be taken when attempting to detect planets in binary systems using eclipse timings. All other period change effects must be taken into account. Regularly sampled, long base lines should be used. 

We attempted to detect a variation in the size of the secondary stars in the systems GK Vir and NN Ser. For both systems the measured radii appear consistent throughout. However, the accuracy of our measurements imply that we may be able to detect changes in the size of the stars due to Applegate's mechanism in the future. We find no evidence for Wilson depressions in either of these systems.  

Inspection of the ULTRACAM light curves shows that the rate of flaring of the secondary stars is different in each of the systems. The data hint that flaring rates depend more on the mass of the secondary star rather than its rotation rate, even though these are all fast rotators. 

\section*{Acknowledgements}

TRM, CMC and BTG acknowledge support from the Science and Technology Facilities Council (STFC) grant number ST/F002599/1. SPL acknowledges the support of an RCUK Fellowship. ULTRACAM, VSD and SPL are supported by STFC grants ST/G003092/1 and PP/E001777/1. EU was supported by Universidad Cat{\'o}lica del Norte, DGIP grant number 220401-10301257. The results presented in this paper are based on observations collected at the European Southern Observatory (La Silla) under programme IDs 073.D-0633 and 079.D-0518 and with the William Herschel Telescope operated on the island of La Palma by the Isaac Newton Group in the Spanish Observatorio del Roque de los Muchachos of the Institutions de Astrofisica de Canarias. This paper makes use of SIMBAD, maintained by the Centre de Donn\'{e}es astronomiques de Strasbourg, the National Aeronautics and Space Administration (NASA) Astrophysics Data System and the USNOFS Image and Catalogue Archive, operated by the United States Naval Observatory, Flagstaff Station.

\bibliographystyle{mn2e}
\bibliography{pcebs}

\section*{Supporting Information}

All the flux calibrated light curves presented in this paper may be found in the online version of this article as supporting information.

\appendix
\label{appen}
\section{ULTRACAM and previous eclipse times}

\begin{table*}
 \centering
  \caption{ULTRACAM eclipse times for DE CVn. The first eclipse (cycle number 0) is the same eclipse as in \citet{besselaar07}, our times are consistent with theirs. All observations were made at the WHT.}
  \label{decvn_new}
  \begin{tabular}{@{}lcccccccl@{}}
  \hline
Cycle  & \emph{u'} eclipse & $R_\mathrm{sec}/a$ & \emph{g'} eclipse & $R_\mathrm{sec}/a$ & \emph{r'/i'/z} eclipse & $R_\mathrm{sec}/a$ & Red    \\
Number &   MJD(UTC)        &                   &   MJD(UTC)        &                   &   MJD(UTC)           &                   & Filter \\
       &   MJD(BTDB)       &                   &   MJD(BTDB)       &                   &   MJD(BTDB)          &                   &        \\
 \hline
   0   & 52784.0518541(18) & 0.18109(3)        & 52784.0518445(13) & 0.18110(2)        & 52784.0518410(68)    & 0.18113(13)       & i      \\
       & 52784.0540495(18) &                   & 52784.0540400(13) &                   & 52784.0540364(68)    &                   &        \\
2801   & 53804.0037989(28) & 0.18110(5)        & 53804.0038091(17) & 0.18120(3)        & 53804.0038027(35)    & 0.18131(7)        & r      \\
       & 53804.0082548(28) &                   & 53804.0082650(17) &                   & 53804.0082586(35)    &                   &        \\
2807   & 53806.1886343(17) & 0.18109(3)        & 53806.1886337(16) & 0.18103(3)        & 53806.1886323(38)    & 0.18091(7)        & r      \\
       & 53806.1931064(17) &                   & 53806.1931057(16) &                   & 53806.1931044(38)    &                   &        \\
2809   & 53806.9168898(76) & 0.18142(13)       & 53806.9169041(33) & 0.18120(6)        & 53806.9168679(90)    & 0.18143(16)       & r      \\
       & 53806.9213660(76) &                   & 53806.9213803(33) &                   & 53806.9213441(90)    &                   &        \\
\hline
\end{tabular}
\end{table*}

\begin{table*}
 \centering
 \begin{minipage}{\columnwidth}
  \centering
  \caption{Previous eclipse times for DE CVn. (1) \citet{robb97}, (2) \citet{besselaar07}, (3) \citet{tas04}}
  \label{decvn_old}
  \begin{tabular}{@{}clcccl@{}}
  \hline
Cycle  & Obs\footnote{UVic: Climenhaga Obs, Victoria, Canada. DAO: Dominion Astrophysical Obs,  Victoria, Canada. MDM: Michigan-Dartmouth-MIT Obs, Arizona, USA. EGE: Ege University Obs, Turkey.}   & Eclipse time & Eclipse time & Uncert  & Ref \\
Number &       & MJD(UTC)         & MJD(BTDB)        & MJD     &     \\
 \hline
-6134  & UVic  & 50550.4180       & 50550.4221       & 0.0016  & (1) \\
-6109  & UVic  & 50559.5212       & 50559.5250       & 0.0016  & (1) \\
-6107  & UVic  & 50560.2500       & 50560.2538       & 0.0020  & (1) \\
-6101  & UVic  & 50562.4344       & 50562.4381       & 0.0022  & (1) \\
-6079  & UVic  & 50570.4471       & 50570.4504       & 0.0014  & (1) \\
-6063  & UVic  & 50576.2725       & 50576.2756       & 0.0014  & (1) \\
-6057  & UVic  & 50578.4584       & 50578.4613       & 0.0006  & (1) \\
-4912  & UVic  & 50995.4012       & 50995.4011       & 0.0018  & (2) \\
-3196  & UVic  & 51620.2591       & 51620.2636       & 0.0015  & (2) \\
-2015  & UVic  & 52050.3108       & 52050.3133       & 0.0016  & (2) \\
-2001  & DAO   & 52055.4089       & 52055.4111       & 0.0007  & (2) \\
-1982  & UVic  & 52062.3280       & 52062.3297       & 0.0007  & (2) \\
-1342  & MDM   & 52295.3761       & 52295.3789       & 0.0004  & (2) \\
-1334  & MDM   & 52298.2891       & 52298.2921       & 0.0001  & (2) \\
-1019  & EGE   & 52412.9940       & 52412.9965       & 0.0004  & (3) \\
 -988  & DAO   & 52424.2828       & 52424.2846       & 0.0006  & (2) \\
 -900  & MDM   & 52456.3288       & 52456.3286       & 0.0004  & (2) \\
 -304  & MDM   & 52673.35212      & 52673.35562      & 0.00014 & (2) \\
 -217  & EGE   & 52705.0322       & 52705.0366       & 0.0003  & (3) \\
 -157  & EGE   & 52726.8800       & 52726.8844       & 0.0004  & (3) \\
   69  & MDM   & 52809.1789       & 52809.1795       & 0.0004  & (2) \\
 2873  & UVic  & 53830.2220       & 53830.2263       & 0.0004  & (2) \\
\hline
\end{tabular}
\end{minipage}
\end{table*}

\begin{table*}
 \centering
  \caption{ULTRACAM eclipse times for GK Vir. All observations were made at the WHT except for cycle number 34054 which was made at the VLT.}
  \label{gkvir_new}
  \begin{tabular}{@{}lcccccccl@{}}
  \hline
Cycle  & \emph{u'} eclipse & $R_\mathrm{sec}/a$ & \emph{g'} eclipse & $R_\mathrm{sec}/a$ & \emph{r'/i'/z'} eclipse & $R_\mathrm{sec}/a$ & Red    \\
Number &   MJD(UTC)        &                   &   MJD(UTC)        &                   &   MJD(UTC)           &                   & Filter \\
       &   MJD(BTDB)       &                   &   MJD(BTDB)       &                   &   MJD(BTDB)          &                   &        \\
 \hline
28666  & 52413.9197824(54) & 0.08481(11)       & 52413.9197825(9)  & 0.084679(18)      & 52413.9197816(19)    & 0.08464(4)        & r      \\
       & 52413.9255714(54) &                   & 52413.9255716(9)  &                   & 52413.9255707(19)    &                   &        \\
29735  & 52782.0095805(19) & 0.08471(4)        & 52782.0095823(9)  & 0.084643(18)      & 52782.0095818(42))   & 0.08486(8)        & i      \\
       & 52782.0152254(19) &                   & 52782.0152272(9)  &                   & 52782.0152267(42)    &                   &        \\
29738  & 52783.0426232(18) & 0.08472(3)        & 52783.0426229(7)  & 0.084630(14)      & 52783.0426147(25))   & 0.08478(5)        & i      \\
       & 52783.0482188(18) &                   & 52783.0482185(7)  &                   & 52783.0482102(25)    &                   &        \\
30746  & 53130.1274746(58) & 0.08489(12)       & 53130.1274669(27) & 0.084753(51)      & 53130.1274719(56))   & 0.08463(10)       & i      \\
       & 53130.1336956(58) &                   & 53130.1336878(27) &                   & 53130.1336928(56)    &                   &        \\
32706  & 53805.0171937(54) & 0.08488(10)       & 53805.0171856(23) & 0.084742(44)      & 53805.0171828(28))   & 0.08466(5)        & r      \\
       & 53805.0221235(54) &                   & 53805.0221154(23) &                   & 53805.0221125(28)    &                   &        \\
32709  & 53806.0501184(26) & 0.08476(5)        & 53806.0501170(12) & 0.084652(25)      & 53806.0501123(18))   & 0.08465(3)        & r      \\
       & 53806.0551144(26) &                   & 53806.0551129(12) &                   & 53806.0551082(18)    &                   &        \\
34054  & 54269.1761114(8)  & 0.08467(2)        & 54269.1761097(3)  & 0.084654(6)       & 54269.1761122(9))    & 0.08468(2)        & i      \\
       & 54269.1800885(8)  &                   & 54269.1800868(3)  &                   & 54269.1800893(9)     &                   &        \\
\hline
\end{tabular}
\end{table*}

\begin{table*}
 \centering
 \begin{minipage}{\columnwidth}
  \centering
  \caption{Previous eclipse times for GK Vir. We have applied a light travel time correction ($\sim 480$ seconds) to the times of \citet{green78} since it appears they did not make this correction. (1) \citet{green78}}
  \label{gkvir_old}
  \begin{tabular}{@{}clcccl@{}}
  \hline
Cycle  & Obs\footnote[2]{Pal: Palomar Obs, California, USA. }   & Eclipse time & Eclipse time & Uncert  & Ref \\
Number &       & MJD(UTC)       & MJD(BTDB)      & MJD     &     \\
 \hline
 -67   & Pal   & 42520.26130    & 42520.26747    & 0.00001 & (1) \\
 -32   & Pal   & 42532.31292    & 42532.31905    & 0.00002 & (1) \\
 -29   & Pal   & 42533.34592    & 42533.35204    & 0.00009 & (1) \\
   0   & Pal   & 42543.33179    & 42543.33769    & 0.00001 & (1) \\
   3   & Pal   & 42544.36482    & 42544.37068    & 0.00001 & (1) \\
 851   & Pal   & 42836.35916    & 42836.36314    & 0.00006 & (1) \\
1966   & Pal   & 43220.28679    & 43220.29202    & 0.00012 & (1) \\
2132   & Pal   & 43277.44522    & 43277.45101    & 0.00006 & (1) \\
2896   & Pal   & 43540.51806    & 43540.51972    & 0.00012 & (1) \\
\hline
\end{tabular}
\end{minipage}
\end{table*}

\begin{table*}
 \centering
  \caption{ULTRACAM eclipse times for NN Ser. Cycle numbers up to 44480 are the same eclipses as in \citet{brinkworth06}. Our mid-eclipse times for these eclipses are all consistent with their results. Cycle numbers 53230 and 53237 were observed at the VLT, all others are from the WHT. The \emph{z'} band photometry during cycle 41782 was too poor quality to determine radii.}
  \label{nnser_new}
  \begin{tabular}{@{}lcccccccl@{}}
  \hline
Cycle  & \emph{u'} eclipse & $R_\mathrm{sec}/a$ & \emph{g'} eclipse & $R_\mathrm{sec}/a$ & \emph{r'/i'/z'} eclipse & $R_\mathrm{sec}/a$ & Red    \\
Number &   MJD(UTC)        &                   &   MJD(UTC)        &                   &   MJD(UTC)           &                   & Filter \\
       &   MJD(BTDB)       &                   &   MJD(BTDB)       &                   &   MJD(BTDB)          &                   &        \\
 \hline
38960  & 52411.9413644(26) & 0.16551(16)       & 52411.9413679(6)  & 0.16501(4)        & 52411.9413694(11)    & 0.16504(6)        & r   \\
       & 52411.9470531(26) &                   & 52411.9470566(6)  &                   & 52411.9470581(11)    &                   &     \\
38961  & 52412.0714493(27) & 0.16520(16)       & 52412.0714501(5)  & 0.16504(3)        & 52412.0714503(12)    & 0.16505(6)        & r   \\
       & 52412.0771374(27) &                   & 52412.0771382(5)  &                   & 52412.0771384(12)    &                   &     \\
38968  & 52412.9819958(46) & 0.16605(29)       & 52412.9819913(9)  & 0.16511(4)        & 52412.9819897(19)    & 0.16535(10)       & r   \\
       & 52412.9877084(46) &                   & 52412.9877039(9)  &                   & 52412.9877023(19)    &                   &     \\
38976  & 52414.0226722(48) & 0.16583(28)       & 52414.0226639(7)  & 0.16502(4)        & 52414.0226632(20)    & 0.16519(12)       & r   \\
       & 52414.0283472(48) &                   & 52414.0283389(7)  &                   & 52414.0283382(20)    &                   &     \\
38984  & 52415.0633093(39) & 0.16554(23)       & 52415.0633145(7)  & 0.16506(4)        & 52415.0633146(22)    & 0.16524(12)       & r   \\
       & 52415.0689752(39) &                   & 52415.0689804(7)  &                   & 52415.0689805(22)    &                   &     \\
41782  & 52779.0274986(33) & 0.16536(19)       & 52779.0275064(15) & 0.16482(8)        & 52779.0274749(179)   & No data           & z   \\
       & 52779.0331625(33) &                   & 52779.0331703(15) &                   & 52779.0331388(179)   &                   &     \\
41798  & 52781.1088076(15) & 0.16536(9)        & 52781.1088073(7)  & 0.16496(4)        & 52781.1088090(22)    & 0.16504(12)       & i   \\
       & 52781.1144526(15) &                   & 52781.1144523(7)  &                   & 52781.1144540(22)    &                   &     \\
41806  & 52782.1494576(14) & 0.16517(8)        & 52782.1494595(8)  & 0.16512(4)        & 52782.1494575(21)    & 0.16476(12)       & i   \\
       & 52782.1550908(14) &                   & 52782.1550927(8)  &                   & 52782.1550907(21)    &                   &     \\
41820  & 52783.9706045(21) & 0.16542(12)       & 52783.9706060(8)  & 0.16498(4)        & 52783.9706063(23)    & 0.16486(12)       & i   \\
       & 52783.9762135(21) &                   & 52783.9762150(8)  &                   & 52783.9762153(23)    &                   &     \\
44472  & 53128.9430722(96) & 0.16575(58)       & 53128.9430788(45) & 0.16512(24)       & 53128.9430789(83)    & 0.16528(43)       & i   \\
       & 53128.9486742(96) &                   & 53128.9486808(45) &                   & 53128.9486809(83)    &                   &     \\
44473  & 53129.0731513(50) & 0.16573(27)       & 53129.0731593(28) & 0.16540(15)       & 53129.0731459(54)    & 0.16579(34)       & i   \\
       & 53129.0787552(50) &                   & 53129.0787632(28) &                   & 53129.0787498(54)    &                   &     \\
44474  & 53129.2032333(33) & 0.16549(18)       & 53129.2032314(17) & 0.16528(9)        & 53129.2032323(40)    & 0.16532(21)       & i   \\
       & 53129.2088389(33) &                   & 53129.2088370(17) &                   & 53129.2088379(40)    &                   &     \\
44480  & 53129.9837007(54) & 0.16558(32)       & 53129.9837076(30) & 0.16544(19)       & 53129.9837008(50)    & 0.16541(25)       & i   \\
       & 53129.9893165(54) &                   & 53129.9893234(30) &                   & 53129.9893166(50)    &                   &     \\
49662  & 53804.0615456(61) & 0.16581(37)       & 53804.0615501(25) & 0.16541(13)       & 53804.0615457(30)    & 0.16531(17)       & r   \\
       & 53804.0644522(61) &                   & 53804.0644567(25) &                   & 53804.0644523(30)    &                   &     \\
49663  & 53804.1916188(26) & 0.16596(15)       & 53804.1916184(12) & 0.16524(7)        & 53804.1916174(19)    & 0.16545(11)       & r   \\
       & 53804.1945354(26) &                   & 53804.1945350(12) &                   & 53804.1945340(19)    &                   &     \\
49671  & 53805.2321817(16) & 0.16525(9)        & 53805.2321827(6)  & 0.16496(3)        & 53805.2321818(10)    & 0.16517(6)        & r   \\
       & 53805.2351771(16) &                   & 53805.2351781(6)  &                   & 53805.2351772(10)    &                   &     \\
53230  & 54268.1854167(17) & 0.16533(10)       & 54268.1854172(6)  & 0.16513(4)        & 54268.1854178(13)    & 0.16481(7)        & i   \\
       & 54268.1903109(17) &                   & 54268.1903114(6)  &                   & 54268.1903120(13)    &                   &     \\
53237  & 54269.0960182(6)  & 0.16506(3)        & 54269.0960181(2)  & 0.16497(1)        & 54269.0960191(8)     & 0.16512(4)        & i   \\
       & 54269.1008714(6)  &                   & 54269.1008713(2)  &                   & 54269.1008723(8)     &                   &     \\
56442  & 54686.0062951(23) & 0.16537(14)       & 54686.0062944(9)  & 0.16517(5)        & 54686.0062964(17)    & 0.16525(10)       & r   \\
       & 54686.0076286(23) &                   & 54686.0076279(9)  &                   & 54686.0076299(17)    &                   &     \\
\hline
\end{tabular}
\end{table*}

\begin{table*}
 \centering
 \begin{minipage}{\columnwidth}
  \centering
  \caption{Other eclipse times for NN Ser. (1) \citet{haefner89}, (2) \citet{wood91}, (3) \citet{pigulski02}, (4) \citet{haefner04}, (5) \citet{qian09}, (6) this paper.}
  \label{nnser_old}
  \begin{tabular}{@{}clcccl@{}}
  \hline
Cycle  & Obs\footnote{Dan: Danish 1.5m telescope, La Silla, Chile. ESO: European Southern Observatory 3.6m telescope, La Silla, Chile. McD: McDonald Observatory, Texas, USA. Cal: Calar Alto Observatory, Spain. VLT: Very Large Telescope, Cerro Paranal, Chile. Wro: Bialk{\'o}w station, Wroc{\l}aw University Observatory, Poland. Yun: Lijiang Station, Yunnan Astronomical Observatory, China. NTT: New Technology Telescope, La Silla, Chile.}   & Eclipse time & Eclipse time & Uncert  & Ref \\
Number &       & MJD(UTC)       & MJD(BTDB)      & MJD      &     \\
 \hline
0      & Dan  & 47344.021      & 47344.025      & 0.005     & (1) \\
2760   & ESO  & 47703.041401   & 47703.045744   & 0.000002  & (4) \\
2761   & ESO  & 47703.171497   & 47703.175833   & 0.000006  & (4) \\
2769   & ESO  & 47704.212182   & 47704.216460   & 0.000003  & (4) \\
2776   & ESO  & 47705.122796   & 47705.127023   & 0.000003  & (4) \\
2777   & ESO  & 47705.252896   & 47705.257115   & 0.000007  & (4) \\
2831   & McD  & 47712.27779    & 47712.28158    & 0.00015   & (2) \\
2839   & McD  & 47713.31850    & 47713.32223    & 0.00015   & (2) \\
7360   & McD  & 48301.41331    & 48301.41420    & 0.00015   & (2) \\
28152  & Cal  & 51006.03704    & 51006.04050    & 0.00020   & (4) \\
30721  & VLT  & 51340.21072    & 51340.21590    & 0.00020   & (4) \\
33233  & Wro  & 51666.97227    & 51666.97790    & 0.00040   & (3) \\
58638  & Yun  & 54971.65784    & 54971.66350    & 0.00008   & (5) \\
58645  & Yun  & 54972.56841    & 54972.57406    & 0.00010   & (5) \\
58684  & Yun  & 54977.64160    & 54977.64718    & 0.00012   & (5) \\
58745  & Yun  & 54985.57668    & 54985.58208    & 0.00012   & (5) \\
58753  & Yun  & 54986.61747    & 54986.62284    & 0.00013   & (5) \\
58796  & NTT  & 54992.2110071  & 54992.2161925  & 0.0000015 & (6) \\
\hline
\end{tabular}
\end{minipage}
\end{table*}

\begin{table*}
 \centering
  \caption{ULTRACAM eclipse times for QS Vir. All observations were made at the WHT.}
  \label{qsvir_new}
  \begin{tabular}{@{}lcccccccl@{}}
  \hline
Cycle  & \emph{u'} eclipse & $R_\mathrm{sec}/a$ & \emph{g'} eclipse & $R_\mathrm{sec}/a$ & \emph{r'/i'/z'} eclipse & $R_\mathrm{sec}/a$ & Red    \\
Number &   MJD(UTC)        &                   &   MJD(UTC)        &                   &   MJD(UTC)           &                   & Filter \\
       &   MJD(BTDB)       &                   &   MJD(BTDB)       &                   &   MJD(BTDB)          &                   &        \\
 \hline
24715  & 52415.1074646(20) & 0.21918(9)        & 52415.1074644(6)  & 0.21928(3)        & 52415.1074622(13)    & 0.21941(7)        & r  \\
       & 52415.1133025(20) &                   & 52415.1133023(6)  &                   & 52415.1133001(13)    &                   &    \\
27135  & 52779.9404381(11) & 0.21935(5)        & 52779.9404403(8)  & 0.21926(4)        & 52779.9404357(37)    & 0.21965(18)       & i  \\
       & 52779.9462791(11) &                   & 52779.9462813(8)  &                   & 52779.9462767(37)    &                   &    \\
27149  & 52782.0511392(32) & 0.21975(15)       & 52782.0511443(18) & 0.21972(10)       & 52782.0511456(40)    & 0.22022(23)       & i  \\
       & 52782.0568779(32) &                   & 52782.0568830(18) &                   & 52782.0568844(40)    &                   &    \\
27162  & 52784.0110903(15) & 0.21925(7)        & 52784.0110936(12) & 0.21931(5)        & 52784.0110852(55)    & 0.22174(38)       & i  \\
       & 52784.0167283(15) &                   & 52784.0167316(12) &                   & 52784.0167232(55)    &                   &    \\
33948  & 53807.0500637(25) & 0.21911(12)       & 53807.0500655(11) & 0.21917(5)        & 53807.0500697(20)    & 0.21928(10)       & r  \\
       & 53807.0553148(25) &                   & 53807.0553166(11) &                   & 53807.0553207(20)    &                   &    \\
\hline
\end{tabular}
\end{table*}

\begin{table*}
 \centering
 \begin{minipage}{\columnwidth}
  \centering
  \caption{Other eclipse times for QS Vir. (1) \citet{odonoghue03}, (2) \citet{kawka02}, (3) this paper, (4) \citet{qian10}.}
  \label{qsvir_old}
  \begin{tabular}{@{}clcccl@{}}
  \hline
Cycle  & Obs\footnote{SAAO: South African Astronomical Observatory, Sutherland, South Africa. MSO: Mount Stromlo Observatory, Canberra, Australia. CBA: Bronberg Observatory, Pretoria, South Africa. ESO: European Southern Observatory 3.6m telescope, La Silla, Chile. Yun: Yunnan Astronomical Observatory, China. OCA: Observatorio Cerro Armazones, Chile.}   & Eclipse time & Eclipse time & Uncert  & Ref \\
Number &       & MJD(UTC)       & MJD(BTDB)      & MJD      &     \\
 \hline
171    & SAAO  & 48714.91450    & 48714.92068    & 0.00001  & (1) \\
172    & SAAO  & 48715.06527    & 48715.07146    & 0.00001  & (1) \\
212    & SAAO  & 48721.09541    & 48721.10174    & 0.00001  & (1) \\
225    & SAAO  & 48723.05520    & 48723.06158    & 0.00001  & (1) \\
535    & SAAO  & 48769.79106    & 48769.79641    & 0.00001  & (1) \\
542    & SAAO  & 48770.84646    & 48770.85174    & 0.00001  & (1) \\
2347   & SAAO  & 49042.96519    & 49042.96923    & 0.00001  & (1) \\
2354   & SAAO  & 49044.02042    & 49044.02454    & 0.00001  & (1) \\
2367   & SAAO  & 49045.98011    & 49045.98439    & 0.00001  & (1) \\
2705   & SAAO  & 49096.93400    & 49096.94046    & 0.00001  & (1) \\
3122   & SAAO  & 49159.80281    & 49159.80638    & 0.00001  & (1) \\
4497   & SAAO  & 49367.09788    & 49367.09801    & 0.00001  & (1) \\
4855   & SAAO  & 49421.06420    & 49421.06921    & 0.00001  & (1) \\
5471   & SAAO  & 49513.93137    & 49513.93584    & 0.00001  & (1) \\
7230   & SAAO  & 49779.11374    & 49779.11826    & 0.00001  & (1) \\
7249   & SAAO  & 49781.97794    & 49781.98267    & 0.00001  & (1) \\
7778   & SAAO  & 49861.72778    & 49861.73339    & 0.00001  & (1) \\
7826   & SAAO  & 49868.96457    & 49868.96976    & 0.00001  & (1) \\
7831   & SAAO  & 49869.71840    & 49869.72354    & 0.00001  & (1) \\
9425   & SAAO  & 50110.02959    & 50110.03102    & 0.00001  & (1) \\
9591   & SAAO  & 50135.05298    & 50135.05677    & 0.00001  & (1) \\
9611   & SAAO  & 50138.06788    & 50138.07193    & 0.00001  & (1) \\
10551  & SAAO  & 50279.78259    & 50279.78400    & 0.00001  & (1) \\
11966  & SAAO  & 50493.10273    & 50493.10590    & 0.00001  & (1) \\
12508  & SAAO  & 50574.81016    & 50574.81650    & 0.00001  & (1) \\
15625  & SAAO  & 51044.72960    & 51044.72777    & 0.00001  & (1) \\
17014  & SAAO  & 51254.12438    & 51254.12992    & 0.00001  & (1) \\
17391  & SAAO  & 51310.95933    & 51310.96554    & 0.00001  & (1) \\
23919  & SAAO  & 52295.10958    & 52295.11040    & 0.00001  & (1) \\
24507  & MSO   & 52383.74916    & 52383.75572    & 0.00001  & (2) \\
24520  & MSO   & 52385.70902    & 52385.71558    & 0.00001  & (2) \\
34742  & CBA   & 53926.754380   & 53926.756325   & 0.000017 & (3) \\
34749  & CBA   & 53927.809767   & 53927.811611   & 0.000017 & (3) \\
34762  & CBA   & 53929.769906   & 53929.771562   & 0.000017 & (3) \\
34795  & CBA   & 53934.745279   & 53934.746452   & 0.000029 & (3) \\
34802  & CBA   & 53935.800740   & 53935.801810   & 0.000034 & (3) \\
34808  & CBA   & 53936.705273   & 53936.706254   & 0.000094 & (3) \\
34868  & CBA   & 53945.751723   & 53945.751821   & 0.000020 & (3) \\
38560  & ESO   & 54502.346790   & 54502.349156   & 0.000008 & (3) \\
38566  & ESO   & 54503.251262   & 54503.253715   & 0.000024 & (3) \\
38573  & ESO   & 54504.306406   & 54504.308961   & 0.000027 & (3) \\
38580  & ESO   & 54505.361651   & 54505.364307   & 0.000011 & (3) \\
41270  & Yun   & 54910.896377   & 54910.902126   & 0.000040 & (4) \\
41296  & Yun   & 54914.815887   & 54914.821826   & 0.000040 & (4) \\
41296  & Yun   & 54914.815978   & 54914.821917   & 0.000040 & (4) \\
41302  & Yun   & 54915.720447   & 54915.726426   & 0.000040 & (4) \\
41495  & Yun   & 54944.816001   & 54944.822564   & 0.000040 & (4) \\
43342  & OCA   & 55223.270376   & 55223.271832   & 0.000022 & (3) \\
43349  & OCA   & 55224.325543   & 55224.327104   & 0.000025 & (3) \\
43362  & OCA   & 55226.285138   & 55226.286894   & 0.000033 & (3) \\
43369  & OCA   & 55227.340393   & 55227.342254   & 0.000028 & (3) \\
43415  & OCA   & 55234.274562   & 55234.277101   & 0.000030 & (3) \\
43422  & OCA   & 55235.329786   & 55235.332426   & 0.000014 & (3) \\
\hline
\end{tabular}
\end{minipage}
\end{table*}

\begin{table*}
 \centering
  \caption{ULTRACAM eclipse times for RR Cae. All observations were made at the VLT.}
  \label{rrcae_new}
  \begin{tabular}{@{}lcccccccl@{}}
  \hline
Cycle  & \emph{u'} eclipse & $R_\mathrm{sec}/a$ & \emph{g'} eclipse & $R_\mathrm{sec}/a$ & \emph{r'/i'/z'} eclipse & $R_\mathrm{sec}/a$ & Red    \\
Number &   MJD(UTC)        &                   &   MJD(UTC)        &                   &   MJD(UTC)           &                   & Filter \\
       &   MJD(BTDB)       &                   &   MJD(BTDB)       &                   &   MJD(BTDB)          &                   &        \\
 \hline
 7173  & 53701.0121015(19) & 0.08632(4)        & 53701.0121006(8)  & 0.08632(2)        & 53701.0121116(34)    & 0.08604(7)        & i  \\
       & 53701.0148245(19) &                   & 53701.0148236(8)  &                   & 53701.0148346(34)    &                   &    \\
 7174  & 53701.3158207(32) & 0.08602(13)       & 53701.3158174(4)  & 0.08628(1)        & 53701.3158130(19)    & 0.08628(4)        & i  \\
       & 53701.3185392(32) &                   & 53701.3185359(4)  &                   & 53701.3185315(19)    &                   &    \\
\hline
\end{tabular}
\end{table*}

\begin{table*}
 \centering
 \begin{minipage}{\columnwidth}
  \centering
  \caption{Previous eclipse times for RR Cae. (1) \citet{krzeminski84}, (2) \citet{maxted07}, (3) \citet{brunch98}}
  \label{rrcae_old}
  \begin{tabular}{@{}clcccl@{}}
  \hline
Cycle  & Obs\footnote{LCO: Las Campanas Observatory, Cerro Las Campanas, Chile. SAAO: South African Astronomical Observatory, Sutherland, South Africa. LNA: Laboratorio Nacional de Astrofisica, Pico dos Dias, Brazil.}   & Eclipse time & Eclipse time & Uncert  & Ref \\
Number &       & MJD(UTC)       & MJD(BTDB)      & MJD      &     \\
 \hline
-18423 & LCO   & 45927.415604   & 45927.416650   & 0.000116 & (1) \\
 -5932 & SAAO  & 49720.976772   & 49720.978524   & 0.000003 & (2) \\
 -5929 & SAAO  & 49721.887953   & 49721.889675   & 0.000003 & (2) \\
 -5916 & SAAO  & 49725.836240   & 49725.837828   & 0.000004 & (2) \\
 -2770 & LNA   & 50681.288255   & 50681.289580   & 0.000116 & (3) \\
 -2760 & LNA   & 50684.325599   & 50684.327030   & 0.000116 & (3) \\
 -2750 & LNA   & 50687.362075   & 50687.363610   & 0.000116 & (3) \\
 -2747 & LNA   & 50688.273134   & 50688.274700   & 0.000116 & (3) \\
 -2747 & LNA   & 50688.273204   & 50688.274770   & 0.000116 & (3) \\
 -2708 & SAAO  & 50700.117255   & 50700.119201   & 0.000004 & (2) \\
 -2544 & SAAO  & 50749.923693   & 50749.926547   & 0.000002 & (2) \\
 -2534 & SAAO  & 50752.960724   & 50752.963587   & 0.000002 & (2) \\
 -2524 & SAAO  & 50755.997756   & 50756.000622   & 0.000002 & (2) \\
 -1572 & SAAO  & 51045.125194   & 51045.126463   & 0.000002 & (2) \\
     1 & SAAO  & 51522.849812   & 51522.852260   & 0.000030 & (2) \\
     5 & SAAO  & 51524.064640   & 51524.067060   & 0.000050 & (2) \\
     5 & SAAO  & 51524.064650   & 51524.067070   & 0.000030 & (2) \\
    18 & SAAO  & 51528.012855   & 51528.015180   & 0.000040 & (2) \\
    31 & SAAO  & 51531.961128   & 51531.963350   & 0.000030 & (2) \\
  5616 & SAAO  & 53228.147153   & 53228.148145   & 0.000002 & (2) \\
\hline
\end{tabular}
\end{minipage}
\end{table*}

\begin{table*}
 \centering
  \caption{ULTRACAM eclipse times for RX J$2130.6+4710$, these are the same eclipses as in \citet{maxted04}, our measured eclipse times are consistent with theirs. The eclipse of cycle number -716 featured a flare on the egress hence we do not determine secondary star radii for this eclipse. All observations were made at the WHT.}
  \label{rxj_new}
  \begin{tabular}{@{}lcccccccl@{}}
  \hline
Cycle  & \emph{u'} eclipse & $R_\mathrm{sec}/a$ & \emph{g'} eclipse & $R_\mathrm{sec}/a$ & \emph{r'/i'/z'} eclipse & $R_\mathrm{sec}/a$ & Red    \\
Number &   MJD(UTC)        &                   &   MJD(UTC)        &                   &   MJD(UTC)           &                   & Filter \\
       &   MJD(BTDB)       &                   &   MJD(BTDB)       &                   &   MJD(BTDB)          &                   &        \\
 \hline
-716   & 52412.1216555(21) & No data           & 52412.1216619(9)  & No data           & 52412.1216707(21)    & No data           & r   \\
       & 52412.1211097(21) &                   & 52412.1211161(9)  &                   & 52412.1211249(21)    &                   &     \\
  -2   & 52784.1407462(13) & 0.12273(2)        & 52784.1407520(9)  & 0.12267(1)        & 52784.1407419(38)    & 0.12271(5)        & i   \\
       & 52784.1405462(13) &                   & 52784.1405519(9)  &                   & 52784.1405418(38)    &                   &     \\
   0   & 52785.1827661(13) & 0.12268(2)        & 52785.1827768(11) & 0.12272(1)        & 52785.1827686(42)    & 0.12262(5)        & i   \\
       & 52785.1826194(13) &                   & 52785.1826302(11) &                   & 52785.1826219(42)    &                   &     \\
\hline
\end{tabular}
\end{table*}

\begin{table*}
 \centering
 \begin{minipage}{\columnwidth}
  \centering
  \caption{Previous eclipse times for RX J$2130.6+4710$. These data are not suitable for long-term period studies. (1) \citet{maxted04}.}
  \label{rxj_old}
  \begin{tabular}{@{}clcccl@{}}
  \hline
Cycle  & Obs\footnote[2]{JKT: Jacobus Kapteyn Telescope, La Palma. INT: Isaac Newton Telescope, La Palma.}   & Eclipse time & Eclipse time & Uncert  & Ref \\
Number &       & MJD(UTC)       & MJD(BTDB)      & MJD      &     \\
 \hline
 -1939 & JKT   & 51774.890168   & 51774.893803   & 0.000018 & (1) \\
 -1937 & JKT   & 51775.932234   & 51775.935891   & 0.000018 & (1) \\
 -1935 & INT   & 51776.974257   & 51776.977936   & 0.000005 & (1) \\
\hline
\end{tabular}
\end{minipage}
\end{table*}

\begin{table*}
 \centering
  \caption{ULTRACAM eclipse time for SDSS $0110+1326$ made at the WHT.}
  \label{sdss0110_new}
  \begin{tabular}{@{}lcccccccl@{}}
  \hline
Cycle  & \emph{u'} eclipse & $R_\mathrm{sec}/a$ & \emph{g'} eclipse & $R_\mathrm{sec}/a$ & \emph{r'/i'/z'} eclipse & $R_\mathrm{sec}/a$ & Red    \\
Number &   MJD(UTC)        &                   &   MJD(UTC)        &                   &   MJD(UTC)           &                   & Filter \\
       &   MJD(BTDB)       &                   &   MJD(BTDB)       &                   &   MJD(BTDB)          &                   &        \\
 \hline
 1203  & 54394.1647900(43) & 0.09460(7)        & 54394.1647932(9)  & 0.09463(2)        & 54394.1648007(34)    & 0.09458(7)        & i  \\
       & 54394.1712500(43) &                   & 54394.1712532(9)  &                   & 54394.1712607(34)    &                   &    \\
\hline
\end{tabular}
\end{table*}

\begin{table*}
 \centering
 \begin{minipage}{\columnwidth}
  \centering
  \caption{Previous eclipse times for SDSS $0110+1326$. (1) \citet{pyrzas09}.}
  \label{sdss0110_old}
  \begin{tabular}{@{}clcccl@{}}
  \hline
Cycle  & Obs\footnote[3]{Cal: Calar Alto Observatory, Spain. Mer: Mercator Telescope, La Palma.}   & Eclipse time & Eclipse time & Uncert  & Ref \\
Number &       & MJD(UTC)       & MJD(BTDB)      & MJD      &     \\
 \hline
    0  & Cal   & 53993.94284    & 53993.94870    & 0.00020  & (1) \\
    3  & Cal   & 53994.94062    & 53994.94653    & 0.00020  & (1) \\
    6  & Cal   & 53995.93897    & 53995.94492    & 0.00020  & (1) \\
 1170  & Mer   & 54383.18633    & 54383.19282    & 0.00020  & (1) \\
\hline
\end{tabular}
\end{minipage}
\end{table*}

\begin{table*}
 \centering
  \caption{ULTRACAM eclipse time for SDSS $0303+0054$. Poor observing conditions during eclipse cycle 3058 led to the loss of data in the \emph{u'} band. All readings were taken at the WHT.}
  \label{sdss0303_new}
  \begin{tabular}{@{}lcccccccl@{}}
  \hline
Cycle  & \emph{u'} eclipse & $R_\mathrm{sec}/a$ & \emph{g'} eclipse & $R_\mathrm{sec}/a$ & \emph{r'/i'/z'} eclipse & $R_\mathrm{sec}/a$ & Red    \\
Number &   MJD(UTC)        &                   &   MJD(UTC)        &                   &   MJD(UTC)           &                   & Filter \\
       &   MJD(BTDB)       &                   &   MJD(BTDB)       &                   &   MJD(BTDB)          &                   &        \\
 \hline
 2968  & 54390.1223164(75) & 0.17333(35)       & 54390.1223205(20) & 0.17411(9)        & 54390.1223114(72)    & 0.17498(30)       &  i   \\
       & 54390.1282892(75) &                   & 54390.1282934(20) &                   & 54390.1282842(72)    &                   &      \\
 2976  & 54391.1977788(64) & 0.17357(54)       & 54391.1977839(18) & 0.17402(9)        & 54391.1977802(57)    & 0.17496(34)       &  i   \\
       & 54391.2037850(64) &                   & 54391.2037900(18) &                   & 54391.2037864(57)    &                   &      \\
 3058  & No data           & No data           & 54402.221443(23)  & 0.17438(97)       & 54402.221379(24)     & 0.1739(11)        &  i   \\
       & No data           &                   & 54402.227684(23)  &                   & 54402.227671(24)     &                   &      \\
\hline
\end{tabular}
\end{table*}

\begin{table*}
 \centering
 \begin{minipage}{\columnwidth}
  \centering
  \caption{Previous eclipse times for SDSS $0303+0054$. (1) \citet{pyrzas09}.}
  \label{sdss0303_old}
  \begin{tabular}{@{}clcccl@{}}
  \hline
Cycle  & Obs\footnote{Cal: Calar Alto Observatory, Spain. Mer: Mercator Telescope, La Palma.}   & Eclipse time & Eclipse time & Uncert  & Ref \\
Number &       & MJD(UTC)       & MJD(BTDB)      & MJD      &     \\
 \hline
    0  & Cal   & 53991.11330    & 53991.11741    & 0.00020  & (1) \\
   14  & Cal   & 53992.99498    & 53992.99923    & 0.00020  & (1) \\
   23  & Cal   & 53994.20495    & 53994.20929    & 0.00020  & (1) \\
   44  & Cal   & 53997.02775    & 53997.03229    & 0.00020  & (1) \\
 2559  & Cal   & 54335.14070    & 54335.14302    & 0.00020  & (1) \\
 2589  & Cal   & 54339.17315    & 54339.17583    & 0.00020  & (1) \\
 2960  & Mer   & 54389.04730    & 54389.05324    & 0.00020  & (1) \\
\hline
\end{tabular}
\end{minipage}
\end{table*}

\label{lastpage}

\end{document}